\documentclass[12pt]{article}
\usepackage[utf8]{inputenc}
\usepackage{amsmath,amsfonts,amssymb,epsfig,fp}
\def\be{\begin{equation}\begin{gathered}}
\def\ee{\end{gathered}\end{equation}}
\def\Sum{\sum\limits}
\def\Prod{\prod\limits}
\def\Int{\int\limits}
\def\half{{\textstyle{1\over2}}}
\def\d{\partial}
\def\pd{\partial}

\def\res{{\rm res}}
\def\bs{\boldsymbol}
\def\mf{\mathfrak}
\def\mc{\mathcal}
\def\Oint{\oint\limits}
\def\Res{\mathrm{\,Res\,}}
\def\stackreb#1#2{\mathrel{\mathop{#2}\limits_{#1}}}
\newcommand{\rf}[1]{(\ref{#1})}

\newtheorem{theorem}{Theorem}

\title{Exact conformal blocks for the W-algebras, twist fields
and isomonodromic deformations}

\author{P. Gavrylenko$^{a,b}$, A. Marshakov$^{c,a}$}
\date{\small
$^a${\it Department of Mathematics and Laboratory\\ of Mathematical
Physics, NRU HSE, Moscow, Russia}\\
$^b${\it Bogolyubov Institute for Theoretical Physics, Kyiv, Ukraine}\\
$^c${\it Theory Department, Lebedev Physics Institute and\\
Institute for Theoretical and Experimental Physics, Moscow, Russia}\\
\vspace{0.2cm}gavrylenko@bitp.kiev.ua,\ \ \ mars@itep.ru
}

\numberwithin{equation}{section}

\begin{document}

\maketitle

\begin{abstract}
\medskip
\noindent
We consider the conformal blocks in the theories with extended conformal W-symmetry for the integer Virasoro central charges. We show that these blocks for the generalized twist fields on sphere can be computed exactly in terms of the free field theory on the covering Riemann surface, even for a non-abelian monodromy group. The generalized twist fields are identified with particular primary fields of the W-algebra, and we propose a straightforward way to compute their W-charges. We demonstrate how these exact conformal blocks can be effectively computed using the technique arisen from the gauge theory/CFT correspondence. We discuss also their direct relation with the isomonodromic tau-function for the quasipermutation monodromy data, which can be an encouraging step on the way of definition of generic conformal blocks for W-algebra using the isomonodromy/CFT correspondence.
\end{abstract}

\section{Introduction}

An interest to conformal field theories (CFT) with extended nonlinear W-symmetry generated by the higher spin holomorphic currents has long history, starting from the original work \cite{ZamW}. These theories resemble many features of ordinary CFT (with only Virasoro symmetry), like free field representation and degenerate fields \cite{FZ,FL}, but it already turns to be impossible to construct in generic situation their conformal blocks \cite{Wblocks} (or the blocks for the algebra of higher spin W-currents) which are the main ingredients in the bootstrap definition of the physical correlation functions.

This interest has been seriously supported in the context of rather nontrivial correspondence between two-dimensional CFT and
four-dimensional supersymmetric gauge theory \cite{LMN,NO,AGT}, where the conformal blocks have to be compared with the Nekrasov instanton partition functions \cite{Nek} producing
in the quasiclassical limit the Seiberg-Witten prepotentials \cite{SW}. This correspondence meets serious difficulties beyond the level of the $SU(2)$ gauge quivers on gauge theory side, i.e. for the higher rank gauge groups, which should correspond to the not yet defined generic blocks of the W-conformal theories. It is already clear, however, that the technique developed in two-dimensional CFT can be applied to four-dimensional gauge theories, and vice versa. Following \cite{KriW,AMtau,Quiver} we are going to demonstrate how it can save efforts for the computation of the exact conformal blocks for the twist fields in theories with W-symmetry.

Even in the Virasoro case generic conformal block is a very nontrivial special function \cite{BPZ}, but there exists two important particular cases where the answer is known almost in explicit form -- the correlation functions containing degenerate fields (which are related to the integrals of hypergeometric type) and the exact Zamolodchikov blocks for a nontrivial (though $c=1$) theory \cite{ZamD,ZamAT,ApiZam}~\footnote{Strictly speaking the CFT-Painlev\'e correspondence \cite{Painleve} gives rise to a collection of new exact conformal blocks, coming from the algebraic solutions of Painlev\'e VI.}.
The first class can be generalized to the case of W-algebras, where similar hypergeometric formulas arise in the case of so-called completely degenerate fields \cite{Litv1}. The algebraic definition still exists when degeneracy is not complete, and in this case the most effective way of computation comes from
use of the gauge theory Nekrasov functions.

Below we are going to study the W-analogs of the Zamolodchikov conformal blocks, which do not belong
to the class of algebraic ones. They can be nevertheless computed exactly, partially using the methods
of gauge theories and corresponding integrable systems. We are going to demonstrate also their direct relations with exactly known isomonodromic $\tau$-functions \cite{SJM}, which confirms therefore their role as an important example
of a generic W-block which can be possible defined (for integer central charges) in terms of corresponding isomonodromic problem \cite{PG}.

The exact conformal blocks of the W-algebras are closely related to the correlation functions of the twist fields, studied long ago in the context of perturbative string theory (see e.g. \cite{Knizhnik,BR,orbi}). However, unlike \cite{ZamAT}, the correlators of the twist fields in these papers were not really expressed through the conformal blocks, and therefore their relation to the W-algebras
remained out of interest, so we are going to fill partially this gap.

The paper is organized as follows. In sect.~\ref{ss:twists} we define the correlators of currents on sphere in presence of the twist fields, and show how they can be computed in terms of free conformal
field theory on the cover. In sect.~\ref{ss:wcharges} we identify the twist fields with the primary fields of the W-algebra and propose a way to extract the values of their quantum numbers from the
previously computed correlation functions of the currents. We also show there that these W-charges have obvious meaning in terms of the eigenvalues of the quasipermutation monodromy matrices. In sect.~\ref{ss:SW} we define the result for the exact conformal block in terms of integrable systems. In particular, we show that the main classical contribution to the result satisfies the well-known Seiberg-Witten (SW) period equation \cite{SW,KriW}, moreover, in this case they can be immediately
solved, which gives the most effective way to express the answer through the period matrix and the prime form on the covering surface. Next, in sect.~\ref{ss:isotau} we discuss the connection of the W-algebra conformal blocks with the $\tau$-function of the isomonodromic problem,
 and show that the W-blocks we have constructed correspond in this context to the $\tau$-function for the case of quasipermutation monodromy data. In sect.~\ref{ss:examples} we construct some explicit examples, and some extra technical information (the recursion procedure we have used for construction of correlators of the higher W-currents, the discussion of their OPE with the stress-tensor, and the computation of the asymptotics of
 the period matrix on the cover and its relation with the structure constants in the expansion of the isomonodromic $\tau$-function) is located in the Appendix.

\section{Twist fields and branched covers \label{ss:twists}}
\subsection{Definition}

We start now with the construction of the conformal blocks of $W({\mf{sl}_N})=W_N$ algebra at integer Virasoro central charges $c=N-1$ following the lines of \cite{ZamAT, Knizhnik, BR}.
It is well-known \cite{FL} that $W_N$ algebra has free-field representation in terms of $N-1$ bosonic fields with the currents $J^a(z) = i\d\phi^a(z)$ satisfying operator product expansion (OPE)
\be
J^a(z)J^b(z')=\frac{K^{ab}}{(z-z')^2}+reg.
\ee
where $K^{ab}$ is the scalar product in the Cartan subalgebra $\mf h\subset \mathfrak{g} = \mf{sl}_N$. For the current $J(z)=\Sum_{a=1}^{N-1} h_a J^a(z)=i\pd\phi(z)$, where $h_a$ is the basis in $\mf{h}$, it is useful to introduce explicit components
\be
J_i(z)=(e_i,J(z)),\ \ \ i=1,\ldots,N
\ee
with $\{ e_i\} $ being the weights of the first fundamental or vector representation, so that
\be
J_i(z)J_j(z')=\frac{(e_i,e_j)}{(z-z')^2}+reg.=\frac{\delta_{ij}-\frac1N}{(z-z')^2}+reg.
\label{OPE}
\ee
All high-spin currents of the $W_N$-algebra at $c=N-1$ are elementary symmetric polynomials of $J_i(z)$ ($\Sum_i J_i(z)=0$), e.g. the first three are
\be
T(z)=-W_2(z)=\frac12:(J(z),J(z)):=\frac12\Sum_i:J_i(z)^2:\\
W(z)=W_3(z)=\Sum_{i<j<k}:J_i(z)J_j(z)J_k(z): =\frac13\Sum_i:J_i(z)^3:\\
W_4(z)=\Sum_{i<j<k<l}:J_i(z)J_j(z)J_k(z)J_l(z): =\frac18:\left(\Sum_i J_i^2(z)\right)^2:-\frac14\Sum_i:J_i^4(z):
\label{generators}
\ee
and the primary fields for the current algebra are exponentials
of $\phi(z)\in \mathfrak{h}$
\be
V_{\bs\theta}(z)=e^{i(\bs\theta,\phi(z))}
\ee
with the corresponding eigenvalues $w_k(\bs\theta)$ of the zero modes of the $W_k(z)$-generators given by symmetric functions of $(e_i,\bs\theta)$.

Now we are going to introduce new fields $\mathcal{O}_s(z)$, which are still primary for all
high-spin currents $\{ W_k(z)\}$, but not for the currents $J_i(z)$. They can be realized as monodromy fields
\be
\gamma_q: J(z)\mathcal{O}_s(q)\mapsto s(J(z))\mathcal{O}_s(q)
\label{continuation}
\ee
for some contours $\gamma_q$ encircling the point $q$ on the base curve,
where $s\in \textsf{W}_{\mf{sl}_N}=S_N$ is an element of the corresponding Weyl group.
The particular cases of this construction were known for the
Abelian monodromy group of the cover \cite{BR, ZamAT},
but even there in the cases with $N>2$ they were not identified with $W_N$ primary fields.

Now we are going to construct the \emph{particular} conformal block (on $\mathbb{P}^1$ with global coordinate $z$), where all monodromy fields can be grouped as $\mathcal{O}_s(q_{2i+1})\mathcal{O}_{s^{-1}}(q_{2i+2})$
at $q_{2i+1}\rightarrow q_{2i+2}$, so that one can take an OPE
\be
\label{ope}
\mathcal{O}_s(z)\mathcal{O}_{s^{-1}}(z')=\Sum_{\bs\theta}
C_{s,\bs\theta}(z-z')^{\Delta(\bs\theta)-2\Delta(s)}\left(V_{\bs\theta}(z')+descendants\right)
\ee
and fix the quantum numbers in the intermediate channels, where there are only the fields with
definite $\mathfrak{h} = u(1)^{N-1}$ charges $\frac1{2\pi i}\oint_z d\zeta J(\zeta)V_{\bs\theta}(z)=\bs\theta\cdot V_{\bs\theta}(z)$. In order to do this consider $\mathcal{G}_0(\mathbf{q})=\mathcal{G}_0(q_1,...,q_{2L})$, together with 1-form $\mathcal{G}_1^i(z|\mathbf{q})dz = \mathcal{G}_1^i(z|q_1,...,q_{2L})dz$ and bidifferential $\mathcal{G}_2^{ij}(z,z'|\mathbf{q})dz dz' = \mathcal{G}_2^{ij}(z,z'|q_1,...,q_{2L})dz dz'$, where
\be
\mathcal{G}_0(q_1,...,q_{2L})=\langle\mc{O}_{s_1}(q_1)\mc{O}_{s_1^{-1}}(q_2)...\mc{O}_{s_L}(q_{2L-1})\mc{O}_{s_L^{-1}}(q_{2L})\rangle\\
\mathcal{G}_1^i(z|q_1,...,q_{2L})=\langle J_i(z)\mc{O}_{s_1}(q_1)
\mc{O}_{s_1^{-1}}(q_2)...\mc{O}_{s_L}(q_{2L-1})\mc{O}_{s_L^{-1}}(q_{2L})\rangle \\
\mathcal{G}_2^{ij}(z,z'|q_1,...,q_{2L})=\langle J_i(z)J_j(z')\mc{O}_{s_1}(q_1)\mc{O}_{s_1^{-1}}(q_2)...
\mc{O}_{s_L}(q_{2L-1})\mc{O}_{s_L^{-1}}(q_{2L})\rangle
\label{correlators}
\ee
which become single-valued on the cover $\pi:\mc{C}\rightarrow \mathbb{P}^1$ with the branch points $q_\alpha$
and corresponding monodromies $s_\alpha$. The indices $i,j$ are just labels
of the sheets of this cover, so the multi-valued differentials \rf{correlators}
on $\mathbb{P}^1$ are now expressed in terms of the single-valued
$\mathcal{G}_1(\xi|q_1,...,q_{2L})d\xi$ and
$\mathcal{G}_2(\xi,\xi'|q_1,...,q_{2L})d\xi d\xi'$ on the covering surface $\mc{C}$:
\be
\mathcal{G}^i_1(z|q_1,...,q_{2L})dz=\mathcal{G}_1(z^i|q_1,...,q_{2L})dz^i\\
\mathcal{G}^{ij}_2(z,z'|q_1,...,q_{2L})dzdz'=
\mathcal{G}_2(z^i,z'^j|q_1,...,q_{2L})dz^idz'^j
\label{sheets}
\ee
where $z^i=\pi^{-1}(z)^i$ is the coordinate at $i$'th preimage of the point $z$, not the power (note that number $i$ is not defined globally due to the presence of monodromies). We should also point out that only local deformations of the positions of the branch points $\{q_\alpha\}$ are allowed, since the global ones -- due to nontrivial monodromies -- can change the global structure of the cover $\pi:\mc{C}\rightarrow \mathbb{P}^1$. This
leads in particular to the fact that in the case of non-abelian monodromy group the positions of the branch
points $\{q_\alpha\}$ cannot play the role of the global coordinates on the
corresponding Hurwitz space.\footnote{Although, sometimes
the Hurwitz space of our interest occurs to be  rational,
and in this case one can choose some global coordinates -- but \emph{not} the positions of the branch points. An explicit example is considered below in sect.\ref{ss:examples}.}

\begin{figure}
\centering
\includegraphics{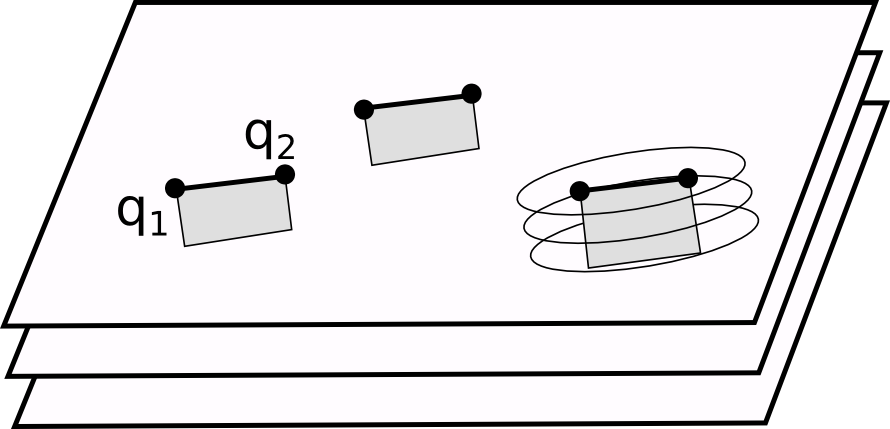}
\caption{Covering Riemann surface $\mathcal{C}$ with simplest cuts between the positions of colliding twist-fields. Sum of the shown cycles of A-type vanishes in $H_1(\mathcal{C})$.}
\label{twists}
\end{figure}
\begin{figure}\centering
\includegraphics{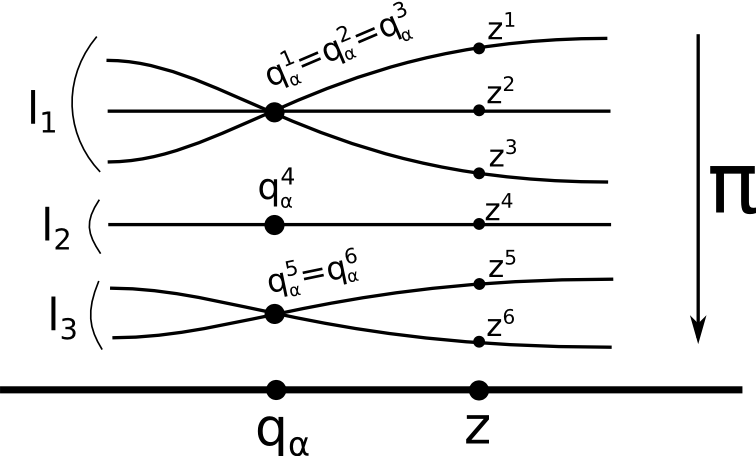}
\caption{Vicinity of a ramification point of a general type.}
\label{branch}
\end{figure}

The picture of the 3-sheeted cover with the most simple branch cuts looks like at fig.\ref{twists},
where we have shown explicitly three (dependent) cycles in $H_1(\mc{C})$ corresponding to the cuts
between the positions of the fields, labeled by mutually inverse permutations.
To understand our notations better we present also at fig.\ref{branch} the picture of the vicinity of the branch-point (of the 6-sheeted cover)
of the cyclic type
$s\sim[3,1,2]$ with several independent permutation cycles.

\subsection{Correlators with the current}

Consider a permutation of the cyclic type $s\sim[l_1,...,l_k]$, which corresponds to the
ramification at $z=q$ (for simplicity we put $q=0$) with $k$ preimages $q^i$, $\pi(q^i)=q$ with
multiplicities $l_i$. The coordinates in the vicinity of these points can be chosen as $\xi_i=z^{1/l_i}$. One can write down a general expression for the expansion of current $J(z)$ on the cover
\be
J(z)=\Sum_{i=1}^k\Sum_{v_i=1}^{l_i-1}\Sum_{n\in\mathbb Z}\frac{a^{(i)}_{n-v_i/l_i}\cdot h_{i,v_i}}{z^{1+n-v_i/l_i}}+\Sum_{j=1}^{k-1}
\Sum_{n\in\mathbb Z}\frac{b^{(j)}_n\cdot H_j}{z^{n+1}}
\ee
where $h_{i,v_i}$ and $H_j$ form the orthogonal basis in $\mf h$ out of the eigenvectors of the permutation $s$, and in coordinates (related to the weights $\{e_i\}$) they have the form
\be
h_{1,v_1}=(1,e^{2\pi i \cdot v_1/l_1},...,e^{2\pi i(l_1-1)\cdot v_1/l_1};0,...,0;...;0,...,0)\\
h_{2,v_2}=(0,...0;1,e^{2\pi i\cdot v_2/l_2},...,e^{2\pi i(l_2-1)\cdot v_2/l_2};0,...,0;...;0,...,0)\\
H_j=(y_j^{(1)},...,y_j^{(1)};y_j^{(2)},...,y_j^{(2)};....;y_j^{(k)},...,y_j^{(k)})\hspace{1cm} \Sum_i l_i{y_j^{(i)}}=0
\ee
with $h_{i,v_i}$, corresponding to non-zero eigenvalues of the permutation cycles $s_i$, while $H_j$ -- to the trivial permutations.

The expansion modes satisfy usual Heisenberg commutation relations $[a_u^{(i)},a_v^{(j)}]=u\delta_{u+v}\delta_{ij}$, $[b_u^{(i)},b_v^{(j)}]=u\delta_{u+v}\delta_{ij}$, up to possible inessential numerical factors which can be extracted from the singularity of the OPE $J(z)J(z')$.
The condition that field $\mc{O}_s(q)$ is primary for the W-currents means in terms of the corresponding state that
\be
a^{(i)}_{u_i}|s\rangle=b^{(j)}_{n}|s\rangle = 0,\ \ \ \ \  u_i>0,\ n>0,\ \forall\ i,j
\ee
and this state is also an eigenvector of the zero modes $b_0^{(j)}$ $\forall\ j$. The corresponding eigenvalues are extra quantum numbers --
 the charges, which have to be included into the definition of the state
$|s\rangle \to |s,{\bf r}\rangle$ (and $\mc{O}_s(q)\to\mc{O}_{s,\mathbf{r}}(q)$) and fixed by expansion of the $\mf h$-valued 1-form $dz J(z)|s\rangle$ at $z\rightarrow 0$, i.e.
\be
\label{Jsing}
\frac{dz}zJ(z)|s,{\bf r}\rangle=\frac{dz}{z}\Sum_{i=1}^N r^ie_i|s,{\bf r}\rangle+reg.
\ee
where $r^1=\ldots=r^{l_1}$, $r^{l_1+1}=\ldots=r^{l_1+l_2}$, etc: the $U(1)$ charges are obviously the same for each point of the cover, they also satisfy the $\mf{sl}_N$ condition
\be
\Sum_{i=1}^N r_\alpha^i=0,\ \ \ \forall\ \alpha
\label{conservation}
\ee
for each branch point $q\in\{q_\alpha\}$. It means that $\mc{G}_1(z)dz$ on the cover $\mathcal{C}$ has only poles with prescribed by \rf{Jsing} singularities,
so one can write
\be
{\mc{G}_1(\xi|{\bf q})d\xi\over \mathcal{G}_0({\bf q})}=\Sum_{\alpha=1}^{2L}d\Omega_{\mathbf{r}_\alpha} + \Sum_{I=1}^g a_I d\omega_I
 =dS
\label{1-form}
\ee
and we shall call this 1-form as the Seiberg-Witten (SW) differential, since its periods over the cycles in $H_1(\mathcal{C})$ play important role in what follows. Here
$\{ d\omega_I\}$, $I=1,\ldots,g$ are the canonically normalized first kind Abelian holomorphic differentials
$$\frac1{2\pi i}\oint_{A_I}d\omega_J=\delta_{IJ}$$ (in slightly unconventional normalization of \cite{Dubrovin} as compare to \cite{Fay,Mumford}), while $$d\Omega_{\mathbf{r}_\alpha} = \sum_{i=1}^N r^i_\alpha d\Omega_{q_{\alpha}^i,p_0}$$ is the third
kind meromorphic
Abelian differential with the simple poles at all preimages of $q_\alpha$  (with the expansion $d\Omega_{\mathbf{r}_\alpha}\stackreb{p^i_\alpha}{=}
l_\alpha^i r^i_\alpha{d\xi_\alpha^i\over\xi_\alpha^i}+reg.$ in corresponding local coordinates) and vanishing A-periods. We denote by $q^i_\alpha=\pi^{-1}(q_\alpha)$, $i=1,\ldots,N$ the preimages on $\mathcal{C}$ of the point $q_\alpha$, with such conventions the point of multiplicity $l_\alpha^i$ has to be counted $l_\alpha^i$ times ($\Res_{p^i_\alpha} d\Omega_{\mathbf{r}_\alpha}=l_\alpha^i r^i_\alpha$).

The A-periods of the differential \rf{1-form}
\be
\label{aper}
a_I = {1\over 2\pi i}\oint_{A_I} dS = {1\over 2\pi i}\oint_{A_I} {d\xi\mc{G}_1(\xi|q)\over \mathcal{G}_0(q)},\ \ \ \ I=1,\ldots,g
\ee
are determined by fixed charges in the intermediate channels due to \rf{ope}.
The number of these constraints
is ensured by the Riemann-Hurwitz formula $\chi(\mathcal{C}) = N\cdot\chi(\mathbb{P}^1) - \# BP$ for the cover
$\pi:\mc{C}\rightarrow \mathbb{P}^1$, or
\be
\label{RH}
g =   \sum_{\alpha=1}^L\sum_{j=1}^{k_\alpha}\left(l_j^\alpha - 1\right) - N + 1 = \sum_{\alpha=1}^{L}\left(N-k_\alpha\right) - N + 1
\ee
where $k_\alpha$ stands for the number of cycles in the permutation $s_\alpha$. One can easily see this in the ``weak-coupling'' regime, when
we can apply \rf{ope} in the limit
$q_{2\alpha-1}\to q_{2\alpha}$, so that
\be
\left.\mathcal{G}_0(q_1,...,q_{2L})\right|_{\bs\theta}=
\langle\left.\mc{O}_{s_1}(q_1)\mc{O}_{s_1^{-1}}(q_2)\right|_{\bs\theta_1}\ldots
\left.\mc{O}_{s_L}(q_{2L-1})\mc{O}_{s_L^{-1}}(q_{2L})\right|_{\bs\theta_L}\rangle\sim
\\
\stackreb{q_{2\alpha-1}\to q_{2\alpha}}{\sim}\ \langle \prod_{\alpha=1}^L V_{\bs\theta_\alpha}(q_{2\alpha})\rangle+\ldots
\ee
and the charge conservation law $\sum_{\alpha=1}^L\bs\theta_\alpha=0$ gives exactly $N-1$ constraints
to the parameters $\{\bs\theta_\alpha\}$, whose total number is $\sum_{\alpha=1}^L\left(N-k_\alpha\right)$, since for each pair of colliding ends of the cut (i.e. $\alpha=1,\ldots,L$) there are
$k_\alpha$ linear relations for the $N$ integrals over the contours, encircling two colliding ramification points, see fig.\ref{twists} (this procedure also gives a way to choose convenient basis in $H_1(\mathcal{C})$ as shown on this picture). For the dual B-periods
of \rf{1-form} one gets
\be
\label{adper}
a^D_I = \oint\limits_{B_I} dS = \mathcal{T}_{IJ}a_J + U_I,\ \ \ \ I=1,\ldots,g
\ee
where the last term can be transformed using the Riemann bilinear relations (RBR) as
\be
\label{U-vector}
U_J = \Sum_{\alpha}\Oint_{B_J}
d\Omega_{\mathbf{r}_\alpha}=\Sum_{\alpha,m}r^m_\alpha A_J(q^m_\alpha),\ \ \ J=1,\ldots,g
\ee
where $A_J(p) = \int_{p_0}^p d\omega_J$ is the Abel map of a point $p\in \mathcal{C}$, and $U_J$ do not
depend on the reference point $p_0\in \mathcal{C}$ due to \rf{conservation}.

\subsection{Stress-tensor and projective connection}

Similarly the 2-differential from \rf{correlators} is fixed by its analytic properties and one can write
\be
{\mathcal{G}_2(p',p|{\bf q})\over\mathcal{G}_0({\bf q})} d\xi_{p'} d\xi_p=dS(p')dS(p)+K(p',p)-\frac1{N}K_0(p',p)
\label{berg}
\ee
where
\be
\label{Faybd}
K(p',p)=d_{\xi_{p'}} d_{\xi_p}\log E(p',p) = \frac{d{\xi_{p'}} d{\xi_p}}{(\xi_{p'}-\xi_p)^2}
 + reg.,
\ \ \ \
\oint_{A_I}K(p',p)=0
\ee
is the canonical meromorphic bidifferential on $\mc{C}$ (the double logarithmic derivative of the prime form, see \cite{Fay}), normalized on vanishing A-periods in each of two variables, while
\be
K_0=\frac{d\pi(\xi)d\pi(\xi')}{(\pi(\xi)-\pi(\xi'))^2}
\ee
is just the pull-back $\pi^*$ of the bidifferential $\frac{dz dz'}{(z-z')^2}$ from $\mathbb{P}^1$.
Formula \rf{berg} is fixed by the following properties: in each of two variables it has almost the same structure as $\mathcal{G}_1(\xi)d\xi$, but with extra singularity on diagonal $p'=p$, which comes from (\ref{OPE}), it also satisfies an obvious condition $\Sum_{i}\mathcal{G}_2^{ij}(z,z')=\Sum_{j}\mathcal{G}_2^{ij}(z,z')=0$

Now one can define \cite{Fay} the projective connection $t_x(p)$ by subtracting the singular part of \rf{Faybd}
\be
\label{tave}
t_x(p) dx^2={1\over 2}\left.\left(K(p',p)-\frac{dx(p') dx(p)}{(x(p')-x(p))^2}\right)\right|_{p'=p}
\ee
It depends on the choice of the local coordinate $x(p)$, and it is easy to check that
\be
\label{Schw}
t_x(p)dx^2-t_\xi(p)d\xi^2=\frac{1}{12}\{\xi,x\}dx^2
\ee
where $\{\xi,x\}= (S\xi)(x) = \frac{\xi_{xxx}}{\xi_x}-\frac32\left(\frac{\xi_{xx}}{\xi_x}\right)^2$ is the Schwarzian derivative.

It is almost obvious that expression \rf{tave} is directly related with the average of the Sugawara stress-tensor
$T(z)$ (\ref{generators}) of conformal field theory (with extended W-symmetry), since normal ordering of free bosonic currents exactly results in subtraction of its singular part. One gets in this way
from \rf{berg} that
\be
\frac{\langle :\frac12J_i(z)J_i(z):\mc{O}_{s_1}(q_1)\mc{O}_{s_1^{-1}}(q_2)\ldots
\mc{O}_{s_L}(q_{2L-1})\mc{O}_{s_L^{-1}}(q_{2L})\rangle}{
\langle\mc{O}_{s_1}(q_1)\mc{O}_{s_1^{-1}}(q_2)...\mc{O}_{s_L}(q_{2L-1})\mc{O}_{s_L^{-1}}(q_{2L})\rangle}=
\\
= t_z(z^i)+{1\over 2}\left(\frac{dS(z^i)}{dz}\right)^2
\ee
where $z=z(p)$ is the global coordinate on $\mathbb{P}^1$, and we have used that after subtraction \rf{tave} one can substitute $K\mapsto 2t_z(p)dz^2$ and $K_0\mapsto 0$, leading to
\be
\langle T(z)\rangle_{\mc O}=\frac{\langle T(z) \mc{O}_{s_1}(q_1)\mc{O}_{s_1^{-1}}(q_2)\ldots
\mc{O}_{s_L}(q_{2L-1})\mc{O}_{s_L^{-1}}(q_{2L})\rangle}{\langle \mc{O}_{s_1}(q_1)\mc{O}_{s_1^{-1}}(q_2)\ldots
\mc{O}_{s_L}(q_{2L-1})\mc{O}_{s_L^{-1}}(q_{2L})\rangle}=
\\
= \Sum_{\pi(p)=z}\left( t_z(p)+
{1\over 2}\left(\frac{dS(p)}{dz}\right)^2\right)
\label{T(z)}
\ee
where sum in the r.h.s. computes the pushforward $\pi_*$, 
appeared here as a result of summation in  \rf{generators}.

\section{W-charges for the twist fields \label{ss:wcharges}}

\subsection{Conformal dimensions for quasi-permutation operators}

Using the OPE with the stress-tensor $T(z)$
\be
\label{bpz}
T(z)\mc{O}_{s,\bf r}(q)=\frac{\Delta(s,{\bf r})\mc{O}_{s,\bf r}(q)}{(z-q)^2}+\frac{\pd_q \mc{O}_{s,\bf r}(q)}{z-q}+reg.
\ee
one can extract from the singularities of \rf{T(z)} the dimensions of the twist fields. Following \cite{BR} we first
notice from \rf{tave} that near the branch point (e.g. at $q=0$) the local coordinate is $\xi_i=z^{1/l_i}$, so that
\be
\label{brasing}
t_z(p)=t_\xi(p)\left(\frac{d\xi}{dz}\right)^2+\frac{1}{12}\{\xi,z\}
=t_\xi(p)z^{2/l_i-2}+\frac{l^2-1}{24l^2}\frac1{z^2}
\ee
The first term in the r.h.s. cannot contain $\frac1{z^2}$-singularity, since $t_\xi(p)$ is regular in local coordinate on the cover $\mc{C}$.
The second source of the second-order pole in \rf{T(z)} comes from the poles of the Seiberg-Witten differential \rf{1-form}, which look as
\be
\label{dSr}
dS \approx r_il_i \frac{d\xi_i}{\xi_i}+reg.= r_i\frac{dz}z+reg.
\ee
Taking them into account together with \rf{brasing} one comes finally to the formula
\be
\Delta(s,{\bf r})=\Sum_{i=1}^k\frac{l_i^2-1}{24l_i}+\Sum_{i=1}^k\frac12 l_ir_i^2
\label{dimension}
\ee
which gives the full conformal dimension for the twist fields with $\mathbf{r}$-charges.

Since we are going to use this formula intensively below, let us illustrate first, how it works
in the first two nontrivial cases:

\begin{itemize}
  \item $N=2$: there are only two possible cyclic types:
\begin{itemize}
\item $s\sim[1,1]$, then $l_1=l_2=1$, $r_1=-r_2=r$, so $\Delta(s,\bs r)=r^2$ is only given by the $r$-charges;
\item $s\sim[2]$, then the only $l_1=2$, the single $r$-charge must vanish, so one just gets here the original Zamolodchikov's twist field with $\Delta(s,\bs r)=\frac1{16}$.
\end{itemize}
  \item $N=3$: here one has three possible cyclic types:
\begin{itemize}
\item $s\sim[1,1,1]$, then $l_1=l_2=l_3=1$, $r_1+r_2+r_3=0$,
$\Delta(s,\bs r)=\frac12\left(r_1^2+r_2^2+r_3^2\right)$
\item $s\sim[2,1]$, then $l_1=2,l_3=1$, $r_1=r_2=r$, $r_3=-2r$,
$\Delta(s,\bs r)=\frac1{16}+3 r^2$
\item $s\sim[3]$, then $l_1=3$, the single $r$-charge again should vanish, so that
the dimension is $\Delta(s,\bs r)=\frac19$.
\end{itemize}

\end{itemize}

\subsection{Quasipermutation matrices}

The hypothesis of the isomonodromy-CFT correspondence \cite{PG} relates the constructed above twist fields to the
quasipermutation monodromies (we return to this issue in more details later). This correspondence relates the $W_N$
charges of the twist fields to the symmetric functions of eigenvalues
of the logarithms of the quasipermutation monodromy matrices
\be
\label{Mtheta}
M_\alpha\sim e^{2\pi i\bs\theta_\alpha},\ \ \ \alpha=1,\ldots,2L\,,
\ee
 being the elements of the semidirect product $S_N\ltimes \left(\mathbb{C}^\times\right)^N$ (here
we consider only the matrices with $\det M_\alpha=1$). An example of the quasipermutation matrix of cyclic type $s\sim[3,2]$ is
\be
M=\begin{pmatrix}0&a_1e^{2\pi i r_1}&0&0&0\\0&0&a_2e^{2\pi i r_1}&0&0\\a_3e^{2\pi i r_1}&0&0&0&0\\0&0&0&0&b_1e^{2\pi ir_2}
\\0&0&0&b_2e^{2\pi ir_2}&0\end{pmatrix}
\ee
where $a_1a_2a_3=1$, $b_1b_2=-1$, $3r_1+2r_2=0$ to get $\det M=1$.
A generic quasipermutation is decomposed into several blocks of the sizes $\{l_i\}$, each of these blocks is given by $$e^{2\pi i r_i}\times
e^{\frac{i\pi}{l_i}\epsilon(l_i)}s_{l_i},\ \ \ \ i=1,\ldots,k$$
where $s_{l_i}$ is the cyclic permutation of length $l_i$, $\epsilon(l)=0$ for $l$-odd
and $\epsilon(l)=1$ for $l$-even. It is easy to check that eigenvalues
of such matrices are
\be
\lambda_{i,v_i}=e^{2\pi i\theta_{i,v_i}}=e^{2\pi i\left(r_i+\frac{v_i}{l_i}\right)},\ \ \ \ i=1,\ldots,k\\
v_i=\frac{1-l_i}2,\frac{1-l_i}2+1\ldots,\frac{l_i-1}2-1,\frac{l_i-1}2
\label{lv}
\ee
According to relation \rf{Mtheta} the conformal dimension of the
corresponding field is
\be
\label{DM}
\Delta(M)=\frac12\Sum\theta_{i,v_i}^2=\frac12\Sum \left(r_i+\frac{v_i}{l_i}\right)^2=\Sum_{i=1}^k\frac{l_i^2-1}{24l_i}+\Sum_{i=1}^k\frac12l_ir_i^2
\ee
where we have used that $\sum v_i=0$ for any fixed $i=1,\ldots,k$, and
\be
\frac{l(l^2-1)}{12} = \left\{\begin{array}{c}
                               \Sum_{-(l-1)/2}^{(l-1)/2}v^2\hspace{1cm}l=2m+1\,\,\, (v\in\mathbb{Z}) \\
                               \Sum_{-(l-1)/2}^{(l-1)/2}v^2\hspace{1cm}l=2m\,\,\, (v\in\mathbb{Z}+\frac12)
                             \end{array}\right.
\ee
for both even or odd $l\in\{l_i\}$. The calculation \rf{DM} for the quasipermutation matrices
reproduces exactly the CFT formula (\ref{dimension}), confirming the correspondence.

\subsection{$W_3$ current}

One can also perform a similar relatively simple check for the first higher $W_3$-current. An obvious generalization of \rf{DM} gives
\be
w_3(M)=\Sum_{a<b<c}(r_a+\frac{v_a}{l_a})(r_b+\frac{v_b}{l_b})(r_c+\frac{v_c}{l_c})=
\frac13\Sum_a(r_a+\frac{v_a}{l_a})^3=\\=
\frac13\Sum_a r_a^3+
\Sum_a r_a\frac{v_a^2}{l_a^2}
=\Sum_{i=1}^k\frac13l_ir_i^3+\Sum_{i=1}^kr_i\frac{l_i^2-1}{12l_i}
\label{w3}
\ee
To extract such formulas from conformal field theory one has to analyze
the multicurrent correlation functions in presence of twist operators
and action of the corresponding modes of the $W_k(z)$ currents. For $W=W_3(z)$, following \rf{correlators} one
can first define
\be
\mathcal{G}_3^{ijk}(z,z',z''|\mathbf{q})dz dz' dz''=
\mathcal{G}_3^{ijk}(z,z',z''|q_1,...,q_{2L})dz dz' dz''=
\\
= \langle J_i(z)J_j(z')J_k(z'')\mc{O}_{s_1}(q_1)\mc{O}_{s_1^{-1}}(q_2)...
\mc{O}_{s_L}(q_{2L-1})\mc{O}_{s_L^{-1}}(q_{2L})\rangle dz dz' dz''
\ee
and write, similarly to \rf{berg}
\be
{\mathcal{G}_3(p'',p',p|{\bf q})\over\mathcal{G}_0({\bf q})} d\xi_{p''}d\xi_{p'} d\xi_p=dS(p'')dS(p')dS(p)+
\\
+ dS(p'')\left(K(p',p)-\frac1{N}K_0(p',p)\right) + dS(p')\left(K(p'',p)-\frac1{N}K_0(p'',p)\right) +
\\
+ dS(p)\left(K(p'',p')-\frac1{N}K_0(p'',p')\right)
\label{berg3}
\ee
where the r.h.s. has appropriate singularities at all diagonals and correct $A$-periods in each of three variables. Extracting singularities and using \rf{generators}, \rf{tave} one can write
\be
\langle W(z)\rangle_{\mc O}=
\Sum_{\pi(p)=z}\left( \frac13 \left(\frac{dS(p)}{dz(p)}\right)^3+2 t_z(p)\frac{dS(p)}{dz(p)} \right)
\ee
It is easy to see that due to \rf{brasing}, \rf{dSr} this formula gives the same result as \rf{w3}.

Formula \rf{lv} also shows, how the charges of the twist fields can be seen within the context of W-algebras. It is important, for example, that for the complete cycle permutation one would get its $W_N$ charges $w_2(\bs\theta), w_3(\bs\theta),\ldots, w_N(\bs\theta)$, where
\be
\bs\theta=\frac{\bs\rho}{N} = \frac1N\left(\frac{N-1}{2},\frac{N-1}{2}-1,\ldots,
\frac{1-N}{2}+1,\frac{1-N}{2}\right)
\ee
i.e. the vector of charges is proportional to the Weyl vector of $\mathfrak{g} = \mf{sl}_N$. Such
fields are non-degenerate from the point of view of the $W_N$ algebra, since for degenerate fields
the charge vector always satisfy the condition $(\bs\theta,\alpha)\in\mathbb{Z}$ for some root $\alpha$. It means that here we are
beyond the algebraically defined conformal blocks, and further investigation of descendants $W_{-1}\mathcal{O}$ etc can shed light on
the structure of generic conformal blocks for the W-algebras. We are going to return to this issue elsewhere.

\subsection{Higher W-currents \label{ss:higher}}

For the higher W-currents ($W_k(z)$ with $k>3$) the situation becomes far more complicated.
We discuss here
briefly only the case of $W_4(z)$, which already gives a hint on what happens in generic situation. An analog of \rf{DM}, \rf{w3} gives for the quasipermutation matrices
\be
w_4(M)=\Sum_{a<b<c<d}(r_a+\frac{v_a}{l_a})(r_b+\frac{v_b}{l_b})(r_c+\frac{v_c}{l_c})(r_d+\frac{v_d}{l_d})
=\frac12 \Delta(M)^2-\frac14 A
\label{w4M}
\ee
with $\Delta(M)$ given by \rf{DM} and
\be
A=\Sum_{a=1}^N\left(r_a+\frac{v_a}{l_a}\right)^4
=\Sum_{i=1}^k l_i r_i^4+6\Sum_{i=1}^k r_i^2\frac{l_i^2-1}{12 l_i}+\Sum_{i=1}^k\frac{(l_i^2-1)(3l_i^2-7)}{240 l_i^3}
\label{AB}
\ee
To get this from CFT one needs just the most singular part of the correlation function
\be
\langle W_4(z)\rangle_{\mc O}(dz)^4\stackreb{z\to q}=w_4\left(\frac{dz}{z-q}\right)^4+\ldots
\label{w4CFT}
\ee
which is a particular case of the current correlators
\be
\mc{R}_{i_1,\ldots i_n}(z_1,\ldots z_n)=\langle:J_{i_1}(z_1),\ldots J_{i_n}(z_n):\rangle_{\mc O}dz_1\ldots dz_n
\label{Rcor}
\ee
and the technique of calculation of such expressions is developed in Appendix~\ref{ap:diag}.

From the definition of the $W_4(z)$ current \rf{generators} it is clear, that
one should take only the most singular parts of the correlation functions of four currents

\begin{picture}(100,40)

\put(20,17){
\put(0,0){$\mc{R}_{iiii}(z,z,z,z)=$}

\multiput(0,0)(80,0){3}{
\multiput(102,12)(0,-20){2}{\circle*{5}}\put(94,12){$i$}\put(94,-9){$i$}
\multiput(122,12)(0,-20){2}{\circle*{5}}\put(126,12){$i$}\put(126,-9){$i$}}
\multiput(143,0)(80,0){2}{$+$}

\put(162,0){$6\,\cdot$}
\put(182,-10){\line(0,1){20}}
\put(242,0){$3\,\cdot$}
\put(262,-10){\line(0,1){20}}
\put(282,-10){\line(0,1){20}}

\put(300,0){$=$}
}

\end{picture}
\be
=dS( z^i)^4+6dS( z^i)^2\hat K_{ii}(z,z)+3\hat K_{ii}(z,z)^2
\label{R1}
\ee
and

\begin{picture}(100,90)

\put(20,67){
\put(0,0){$\mc{R}_{iijj}(z,z,z,z)=$}

\multiput(0,0)(80,0){3}{
\multiput(102,12)(0,-20){2}{\circle*{5}}\put(94,12){$i$}\put(94,-9){$i$}
\multiput(122,12)(0,-20){2}{\circle*{5}}\put(126,12){$j$}\put(126,-9){$j$}}
\multiput(143,0)(80,0){3}{$+$}

\put(182,-10){\line(0,1){20}}
\put(282,-10){\line(0,1){20}}
}

\put(100,17){

\multiput(-80,0)(80,0){3}{
\multiput(102,12)(0,-20){2}{\circle*{5}}\put(94,12){$i$}\put(94,-9){$i$}
\multiput(122,12)(0,-20){2}{\circle*{5}}\put(126,12){$j$}\put(126,-9){$j$}}
\multiput(-17,0)(80,0){3}{$+$}

\put(2,0){$4\,\cdot$}
\put(22,12){\line(1,0){20}}

\put(102,-10){\line(0,1){20}}
\put(122,-10){\line(0,1){20}}
\put(162,0){$2\,\cdot$}
\put(182,-8){\line(1,0){20}}
\put(182,12){\line(1,0){20}}

\put(223,0){$=$}
}

\end{picture}
\be
=dS( z^i)^2dS( z^j)^2+\hat K_{ii}(z,z)dS( z^j)^2+\hat K_{jj}(z,z)dS( z^i)^2+\\+
4\hat K_{ij}(z,z)dS( z^i)dS( z^j)+\hat K_{ii}(z,z)\hat K_{jj}(z,z)+2\hat K_{ij}(z,z)^2
\label{R2}
\ee
taken at the coinciding values of all arguments. It means, that one has to substitute
\be
dS( z^i)=r_i\frac{dz}{z}+\ldots
\label{VEV}
\ee
(we again put here $q=0$ for simplicity)
and do the same for the propagator $\hat K_{ij}(z_1,z_2)=K( z_1^i, z_2^j)-\delta_{ij}K_0(z_1,z_2)$
(see Appendix~\ref{ap:diag} for details), i.e. to substitute into \rf{R1}, \rf{R2}
\be
\hat K_{ii}(z,z)=\left.\frac{dz^{1/l}d\tilde{z}^{1/l}}{(z^{1/l}-\tilde{z}^{1/l})^2}-\frac{dz d\tilde{z}}{(z-\tilde{z})^2}\right|_{z\to \tilde{z}}+\ldots =
\frac{l^2-1}{12l^2}\frac{(dz)^2}{z^2}+\ldots \\
\hat K_{ij}(z,z)=\frac{\zeta^i dz^{1/l} \zeta^j dz^{1/l}}{(\zeta^i z^{1/l}-\zeta^j z^{1/l})^2}+\ldots=\frac1{l^2}
\frac{\zeta^{i-j}}{(1-\zeta^{i-j})^2}\frac{(dz)^2}{z^2}+\ldots \label{propagator}
\ee
where $\zeta=\exp\left(\frac{2\pi i}l\right)$. In order to compute
$-\frac14\Sum_i \mc{R}_{iiii}(z,z,z,z)+\frac18\Sum_{i,j}\mc{R}_{iijj}(z,z,z,z)$
it is useful to move the term $K_{ij}(z,z)^2$ from the second expression to the first one, which gives
\be
\Sum_i dS( z^i)^4 + 6 \Sum_i dS( z^i)^2 \hat K_{ii}(z,z) + 3\Sum_i \hat K_{ii}(z,z)^2-
\\
-\Sum_{ij}\hat K_{ij}(z,z)^2\stackreb{\rf{VEV},\rf{propagator}}{\rightarrow} A\left(\frac{dz}z\right)^4
\label{AA}
\ee
while the rest from \rf{R2} gives rise to
\be
\left(\Sum_i dS( z^i)^2+\Sum_i \hat K_{ii}(z,z)\right)^2+4\Sum_{ij}\hat K_{ij}(z,z)dS( z^i)dS( z^j)
\stackreb{\rf{VEV},\rf{propagator}}{\rightarrow} 4\Delta_N^2\left(\frac{dz}z\right)^4
\label{BB}
\ee
after using \rf{VEV}, \rf{propagator} and several nice formulas like
\be
\frac1l\Sum_{j=1}^{l-1}\frac{\zeta^j}{(1-\zeta^j)^2}=\frac1l\Sum_{j=1}^{l-1}\frac{e^{2\pi ij/l}}{(1-e^{2\pi ij/l})^2}= -\Sum_{v=(1-l)/2}^{(l-1)/2}\frac{v^2}{l^2}\\
\frac1{l^3}\Sum_{j=1}^{l-1}\frac{\zeta^{2j}}{(1-\zeta^j)^4}=
\frac1{l^3}\Sum_{j=1}^{l-1}\frac{e^{4\pi ij/l}}{(1-e^{2\pi ij/l})^4}=
\frac2l\left(\Sum_{v=(1-l)/2}^{(l-1)/2}\frac{v^2}{l^2}\right)^2-
\Sum_{v=(1-l)/2}^{(l-1)/2}\frac{v^4}{l^4}
\ee
Here the sum over the roots of unity can be performed using the contour integral
\be
\Sum_{j=1}^{l-1}\frac{\zeta^{jm}}{(1-\zeta^j)^{2m}}=\frac1{2\pi i}\oint\limits_{z\neq 1}d\log\frac{z^l-1}{z-1}\cdot\frac{z^m}{(1-z)^{2m}}=
\\
= \Res_{z=1}d\log\frac{z-1}{z^l-1}\cdot\frac{z^m}{(1-z)^{2m}}
\ee
and the result indeed allows to identify the coefficients at maximal singularities in \rf{AA}, \rf{BB}
with the expressions \rf{AB}. It means that the conformal charge \rf{w4CFT} of the twist field
indeed coincide with the corresponding symmetric function \rf{w4M} of the eigenvalues of the permutation matrix, but it comes here already from a nontrivial computation.

It is known from long ago that already a definition of the higher W-currents is a nontrivial issue (see e.g. \cite{FZ,FL,Bilal,MM,FLitv}). Here it was important to consider the particular (normally ordered) symmetric function of the currents \rf{generators}, since, for example, another natural choice $\Sum_i:J^4_i(z):$ is even not contained in the algebra generated by $T(z)$, $W_3(z)$ and $W_4(z)$.
However, the so defined $W_4(z)$-current is not a primary field of conformal algebra, we discuss this issue in Appendix~\ref{ap:primarity}.

\section{Conformal blocks and $\tau$-functions \label{ss:SW}}

Consider now the next singular term from the OPE \rf{bpz}, which immediately allows to extract from \rf{T(z)} the accessory parameters
\be
\label{dlogG}
\frac{\pd}{\pd{q_\alpha}}\log\mathcal{G}_0(q_1,...,q_{2L})=\Sum_{\pi(q_\alpha^i)=q_\alpha}\Res_{q_\alpha^i} t_zdz+
\frac12\Sum_{\pi(q_\alpha^i)=q_\alpha}\Res_{q_\alpha^i}\frac{\left(dS\right)^2}{dz}
\ee
Computing residues in the r.h.s. one gets the set of differential equations ($\alpha=1,\ldots,2L$), which define the correlation function of the twist fields $\mathcal{G}_0(q_1,...,q_{2L})$ itself. A non-trivial statement
\cite{KriW,Quiver,KotKor1,KotKor2} is that
these equations are compatible, moreover \rf{dlogG} defines actually two different functions $\tau_{SW}(\mathbf{q})$ and $\tau_B(\mathbf{q})$, where
\be
\label{SW}
\frac{\pd}{\pd{q_\alpha}}\log\tau_{SW}(q_1,...,q_{2L})=
\frac12\Sum_{\pi(q_\alpha^i)=q_\alpha}\Res_{q_\alpha^i}\frac{\left(dS\right)^2}{dz}
\ee
and
\be
\label{KK}
\frac{\pd}{\pd{q_\alpha}}\log\tau_B(q_1,...,q_{2L})=\Sum_{\pi(q_\alpha^i)=q_\alpha}\Res_{q_\alpha^i} t_zdz
\ee
so that $\mathcal{G}_0({\bf q})=\tau_{SW}({\bf q})\cdot\tau_B({\bf q})$, and the claim of  \cite{AMtau,KotKor1,KotKor2}  is that both them are well-defined separately.

\subsection{Seiberg-Witten integrable system}

Let us concentrate attention on $\tau_{SW} = \tau_{SW}(\mathbf{a},\mathbf{q})$ or the Seiberg-Witten prepotential $\mathcal{F} = \log\tau_{SW}$, which is the main contribution to conformal block, and the only one, which depends on the charges in the intermediate channel. According to \cite{AMtau,Quiver}
$\mathcal{F}(\mathbf{a},\mathbf{q})$, up to some possible only $a$-dependent term, satisfies also another set of equations
\be
\label{Bper}
\frac{\pd}{\pd{a_I}}\log\tau_{SW}=a^D_I,\ \ \ \ I=1,\ldots,g
\ee
where the dual periods $a_I^D$ are defined in \rf{adper}. The total system of equations \rf{SW}, \rf{Bper} is
also integrable \cite{KriW,AMtau,Quiver} due to the Riemann bilinear relations. Moreover, in our case
this system of equations can be easily solved due to
\begin{theorem} Function
\be
\label{t1}
\log\tau_{SW}=\half\Sum_{I,J}a_I\mc{T}_{IJ}a_J+\sum_I a_IU_I+\half Q({\bf r})
\ee
solves the system \rf{SW}, iff $Q(\bs r)$ solves the system $\frac{\pd Q(\bs r)}{\pd q_\alpha}=\Sum_{\pi(q_\alpha^i)=q_\alpha}
\Res_{q_\alpha^i}\frac{(d\Omega)^2}{dz} $ for $\alpha=1,\ldots,2L$, $d\Omega=\Sum_\alpha d\Omega_{\mathbf{r}_\alpha}$ and other ingredients in the r.h.s. are given by \rf{aper}, \rf{U-vector} and the period matrix of $\mathcal{C}$.
\end{theorem}
One can check this statement explicitly, using the definitions \rf{1-form} and \rf{U-vector}
\be
\Sum_{\pi(q_\alpha^i)=q_\alpha}\Res_{q_\alpha^i}\frac{d\omega_Id\omega_J}{dz} = -
\Sum_{\pi(q_\alpha^i)=q_\alpha}\Res_{q_\alpha^i}\frac{\d\omega_I}{\d q_\alpha}d\omega_J = -
\oint_{\d\mc C}\frac{\d\omega_I}{\d q_\alpha}d\omega_J =
\\
= {\d\over\d q_\alpha}\oint_{B_I}d\omega_J = {\d\mathcal{T}_{IJ}\over\d q_\alpha}\,,
\label{SWR1}
\ee
where we have first applied the formula $\frac{\pd\omega_I}{\pd q_\alpha}=-\frac{d\omega_I}{dz}+hol.$ and then the RBR. Similarly, for the second term:
\be
\Sum_{\pi(q_\alpha^i)=q_\alpha}\Res_{q_\alpha^i}\frac{d\omega_I
d\Omega_{\mathbf{r}_\alpha}}{dz} = - \Sum_{\pi(q_\alpha^i)=q_\alpha} \Res_{q_\alpha^i}\frac{
\d\Omega_{\mathbf{r}_\alpha}}{\d q_\alpha}d\omega_I = -
\oint_{\d\mc C}\frac{
\d\Omega_{\mathbf{r}_\alpha}}{\d q_\alpha}d\omega_I =
\\
= {\d\over\d q_\alpha}\oint_{B_I}d\Omega_{\mathbf{r}_\alpha} =
{\d U_I \over \d q_\alpha}\,,
\label{SWR2}
\ee
while the last term $Q({\bf r})$, vanishing after taking the $a$-derivatives, should be computed separately,
and the proof will be completed in next section.

\subsection{Quadratic form of $\mathbf{r}$-charges}

In the limit $a_I=0$ equation \rf{SW} gives us the formula
\be
\label{SWQ}
\frac{\pd}{\pd q_\alpha}Q({\bf r})=\Sum_{q^i_\alpha\in \pi^{-1}(q_\alpha)}\Res_{q^i_\alpha}\frac{d\Omega^2}{dz}
\ee
where
\be
d\Omega=\Sum_\alpha d\Omega_{\mathbf{r}_\alpha} = \Sum_{\alpha,i}r_\alpha^id\Omega_{q_\alpha^i,p_0}
\ee
\begin{theorem}
Regularized expression for $Q({\bf r})$
\be
Q({\bf r})_{\vec\epsilon}=\Sum_{\alpha,i} r_\alpha^i\int_{\tilde p_0}^{(q_\alpha^i)_{\epsilon_\alpha}}d\Omega
\label{Qreg}
\ee
satisfies \rf{SWQ} in the limit $\epsilon\to0$
\end{theorem}
{\bf Proof:} It is useful to introduce the differential with shifted poles
\be
d\Omega_{\vec\epsilon}=\Sum_{\alpha,i}r_\alpha^id\Omega_{(q_\alpha^i)_{\epsilon_\alpha},\tilde p_0}\\
\ee
Note that due to conditions \rf{conservation} nothing depends on the reference points $p_0,\tilde p_0$.
The regularized points $(q_i^\alpha)_{\epsilon_\alpha}$ are defined in such a way that
\be
z\left((q_i^\alpha)_{\epsilon_\alpha}\right)= z\left(q_i^\alpha\right)-\epsilon_\alpha = q_\alpha-\epsilon_\alpha
\ee
and this is the only place where the coordinate $z$ on $\mathbb{P}^1$ enters the definition of $Q({\bf r})$. All other parts of
$\tau_{SW}$ do not depend explicitly on the choice of the coordinate $z$ because
they are given by the periods of some meromorphic differentials on the covering
curve.
Expression \rf{Qreg} can now be rewritten equivalently
\begin{figure}
\centering
\includegraphics[width=4cm]{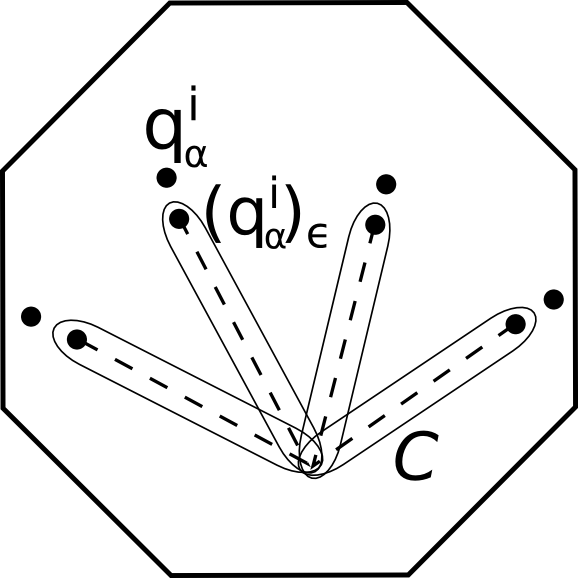}
\caption{Integration path for $Q({\bf r})_{\vec\epsilon}$}
\label{integration}
\end{figure}

\be
Q({\bf r})_{\vec\epsilon}=-\frac1{2\pi i}\oint_{C}\Omega_{\vec\epsilon}\,d\Omega
\ee
where contour $C$ (see fig.\ref{integration}) encircles the branch-cuts of $\Omega_{\vec\epsilon}$, while the poles of $d\Omega$ are left outside. Taking the derivatives one gets
\be
\frac{\pd}{\pd q_\alpha}Q({\bf r})_{\vec\epsilon}=\frac1{2\pi i}\oint_{C}\left[\frac{\pd\Omega}{\pd q_\alpha}
\,d\Omega_{\vec\epsilon}-\frac{\pd\Omega_{\vec\epsilon}}{\pd q_\alpha}
\,d\Omega\right]
\ee
where each of the terms in r.h.s. contains only the poles at the points $q_\alpha^i$ and $(q_\alpha^i)_{\epsilon_\alpha}$ correspondingly. One can therefore shrink the contour of integration in the first term onto the points $q_\alpha^i$ (up to the integration over the boundary of cut Riemann surface, which vanishes due to the Riemann bilinear
relations for the differentials with vanishing $A$-periods), and in the second -- to the points $(q_\alpha^i)_{\epsilon_\alpha}$, hence
\be
\frac{\pd}{\pd q_\alpha}Q({\bf r})_{\vec\epsilon}=-\Sum_i\Res_{q_\alpha^i}\frac{\pd\Omega}{\pd q_\alpha}
\,d\Omega_{\vec\epsilon}-\Sum_i\Res_{(q_\alpha^i)_{\epsilon_\alpha}}\frac{\pd\Omega_{\vec\epsilon}}{\pd q_\alpha}
\,d\Omega
\ee
Near the point $p_\alpha^i$ one can choose the local coordinate $\xi$ such that $z=q_\alpha+\xi^l$,
so that expansion of Abelian integrals can be written as
\be
\Omega=r_i\log(z-q_\alpha)+c_0({\bf q})+c_1({\bf q})(z-q_\alpha)^{1/l}+c_2({\bf q})(z-q_\alpha)^{2/l}+\ldots\\
\Omega_{\vec\epsilon}=\tilde c_0({\bf q})+\tilde c_1({\bf q})(z-q_\alpha)^{1/l}+\tilde
c_2({\bf q})(z-q_\alpha)^{2/l}+\ldots\\
\ee
giving rise to
\be
\frac{\pd\Omega}{\pd q_\alpha}=-\frac{d\Omega}{dz}+\frac{\pd c_0({\bf q})}{\pd q_\alpha}+O\left((z-q_\alpha)^{1/l}\right)\\
\frac{\pd\Omega_{\vec\epsilon}}{\pd q_\alpha}=-\frac{d\Omega}{dz}+\frac{\pd\tilde c_0({\bf q})}{\pd q_\alpha}+O\left((z-q_\alpha-
\epsilon_\alpha)^{1/l}\right)
\ee
Since the differential $d\Omega_{\vec\epsilon}$ is regular near $z=q_\alpha$, one can ignore the regular part when computing the residues:
\be
\frac{\pd}{\pd q_\alpha}Q({\bf r})_{\vec\epsilon}=\Sum_i\Res_{q_\alpha^i}\frac{d\Omega}{dz}
\,d\Omega_{\vec\epsilon}+\Sum_i\Res_{(q_\alpha^i)_{\epsilon_\alpha}}\frac{d\Omega_{\vec\epsilon}}{dz}
\,d\Omega =
\\
=\Sum_i\frac1{2\pi i}\oint_{q_\alpha^i,(q_\alpha^i)_{\epsilon_\alpha}}
\frac{d\Omega d\Omega_{\vec\epsilon}}{dz}
\label{resQe}
\ee
The r.h.s. of this formula has a limit when $\epsilon_\alpha\to0$, so extracting the singular part
from $Q({\bf r})_{\vec\epsilon}$ (easily found from the explicit formula below)
\be
Q({\bf r})= Q({\bf r})_{\vec\epsilon}-2\Sum\Delta_\alpha\log\epsilon_\alpha
\ee
one gets from \rf{resQe} exactly the formula \rf{SWQ}.
This also completes (together with \rf{SWR1}, \rf{SWR2}) the proof of \rf{t1}.

\hfill $\Box$

Using the integration formula for the third kind Abelian differentials \cite{Fay}
$$\int_a^bd\Omega_{c,d} =
\log\frac{E(c,b)E(d,a)}{E(c,a)E(d,b)}$$
one gets from \rf{Qreg} an explicit expression
\be
Q({\bf r})_{\vec\epsilon}=\Sum_{\alpha,i,\beta,j}r_\alpha^ir_\beta^j\log \frac{E((q_\alpha^i)_{\epsilon_\alpha},q_\beta^j)E(\tilde p_0,p_0)}{ E((q_\alpha^i)_{\epsilon_\alpha},p_0)E(\tilde p_0,q_\beta^j)}=
\Sum_{\alpha,i,\beta,j}r_\alpha^ir_\beta^j \log E((q_\alpha^i)_{\epsilon_\alpha},q_\beta^j)=\\=\Sum_{q_\alpha^i\neq q_\beta^j}
r_\alpha^ir_\beta^j\log E(q_\alpha^i,q_\beta^j)+\Sum_{\alpha,i}(r_\alpha^i)^2 l_\alpha^i\log E((q_\alpha^i)_{\epsilon_\alpha},q_\alpha^i)
\ee
The first term in the r.h.s. is regular, while for the second one can use
\be
E((q_\alpha^i)_{\epsilon_\alpha},q_\alpha^i)=
\frac{(z-q_\alpha+\epsilon_\alpha)^{1/l_\alpha^i}-(z-q_\alpha)^{1/l_\alpha^i}}{
\sqrt{d(z-q_\alpha+\epsilon_\alpha)^{1/l_\alpha^i}
d(z-q_\alpha)^{1/l_\alpha^i}}}\approx
\left.\frac{\epsilon_\alpha^{1/l_\alpha^i}}{d\left[(z-q_\alpha)^{1/l_\alpha^i}\right]}\right|_{z\to q_\alpha}
\ee
Therefore
\be
Q({\bf r})=\Sum_{q_\alpha^i\neq q_\beta^j}r_\alpha^ir_\beta^j\log E(q_\alpha^i,q_\beta^j)-\left.\Sum_{\alpha,i}(r_\alpha^i)^2
l_\alpha^i\log d[(z-q_\alpha)^{1/l}]\right|_{z\to q_\alpha}
\ee
Substituting expression of the prime form
\be
E(p,p')=\frac{\Theta_*(A(p)-A(p'))}{h_*(p)h_*(p')}
\ee
in terms of some odd theta-function $\Theta_*$, the already defined above Abel map $A(p)$, and
holomorphic differential
\be
h^2_*(p)=\Sum_I\frac{\pd\Theta_*(0)}{\pd A_I}d\omega_I(p)
\ee
one can write more explicitly
\be
Q({\bf r})=\Sum_{q_\alpha^i\neq q_\beta^j}r_\alpha^ir_\beta^j \log\Theta_*(A(q_\alpha^i)-A(q_\beta^j))-\left.\Sum_{q_\alpha^i}
(r_\alpha^i)^2l_\alpha^i\log\frac{d(z(q)-q_\alpha)^{1/l_\alpha^i}}{h^2_*(q)}\right|_{q=q_\alpha^i}
\label{Qfin}
\ee
If cover $\mathcal{C}$ has zero genus $g(\mathcal{C})=0$ itself, the prime form is just $E(\xi,\xi')=\frac{\xi-\xi'}{\sqrt{d\xi}\sqrt{d\xi'}}$ in terms of the globally defined
coordinate $\xi$, and formula \rf{Qfin} acquires the form
\be
Q({\bf r})=\Sum_{p_\alpha^i\neq p_\beta^j}r_\alpha^ir_\beta^j \log(\xi_\alpha^i-\xi_\beta^j)-\left.\Sum_{\xi_\alpha^i}
(r_\alpha^i)^2l_\alpha^i\log\frac{d(z(\xi)-q_\alpha)^{1/l_\alpha^i}}{d\xi}\right|_{\xi=\xi_\alpha^i}
\label{Qfin0}
\ee
Below we are going to apply this formula to explicit calculation of a particular example for a genus zero cover, but with a non-abelian monodromy group. The result of the computation clearly shows that $\tau$-function \rf{Qfin0} cannot be expressed already in such case as a function of positions of the ramification points $z=q_\alpha$ on $\mathbb{P}^1$, which means that the corresponding formula for $Q(\bs r)$ from \cite{Korotkin} can be applied only in the case of Abelian monodromy group.

\subsection{Bergman $\tau$-function}

The Bergman $\tau$-function, was studied extensively for the different cases
\cite{Knizhnik,BR,ZamAT} from early days of string theory, mostly using the technique of free conformal theory. Modern results and formalism for
this object can be found in \cite{KotKor1,KotKor2}. Already from its definition \rf{KK} $\tau_B$ can be identified
with the variation w.r.t. moduli of the complex structure of the one-loop effective action in the free field theory on the cover.

We are not going to present here an explicit formula for the general Bergman $\tau$-function, it can be found in \cite[formula 1.7]{KotKor2}. We would like only to point out, that for our purposes of studying the conformal blocks this is
the less interesting part, since it does not depends on quantum numbers of the intermediate channels (it means
in particular, that it can be computed just in free field theory). Below in sect.~\ref{ss:examples} we present the result of its direct computation in the simplest case with non-abelian monodromy group. The result shows that it arises just as some quasiclassical renormalization of the term \rf{Qfin0} in the classical part.

However, as for the SW tau-function, the definition \rf{KK} is easily seen to be consistent. Taking one more derivative one gets from this formula
\be
\label{tBc}
\frac{\pd^2 \log\tau_B(\bs q)}{\pd q_\alpha\pd q_\beta}=\frac{\pd}{\pd q_\beta}
\Sum_{\pi(p)=q_\alpha}\underset{p}{\Res}\frac1{dz(p)}\lim_{p'\to p}\left(K(p',p)-\frac{dz(p') dz(p)}{(z(p')-z(p))^2}\right)=\\
=\Sum_{\pi(p)=q_\alpha}\underset{p}{\Res}\frac1{dz(p)}\lim_{p'\to p}\frac{\pd K(p',p)}{\pd q_\beta}=
\Sum_{\pi(p)=q_\alpha}\underset{p}{\Res}\frac1{dz(p)}\times
\\
\times\lim_{p'\to p}
\Sum_{\pi(p'') = q_\beta}\underset{P}{\Res}\frac{K(p',p'')K(p,p'')}{dz(p'')}=
\Sum_{\genfrac{}{}{0pt}{}{\pi(p)=q_\alpha}{\pi(p'')=q_\beta}}
\underset{p,p''}{\Res}\frac{K(p,p'')^2}{dz(p)d(p'')}\,,
\ee
where we have used the Rauch variational formula \cite[formula 3.21]{Fay1} for the canonical meromorphic bidifferential,  computed in the points $p$ and $p'$
with fixed projections
\be
\label{rauch}
\frac{\pd K(p',p)}{\pd q_\beta}=\Sum_{\pi(P) = q_\beta}\underset{P}{\Res}\frac{K(p',P)K(p,P)}{dz(P)}
\ee
so that the expression in r.h.s. of \rf{tBc} is symmetric w.r.t. $\alpha\leftrightarrow\beta$.

This is certainly a well-known fact, but we would like just to point out here, that the Rauch formula \rf{rauch}, which ensures integrability of \rf{KK} can be easily derived itself from the Wick theorem, using the technique, developed in sect.~\ref{ss:twists} and Appendix~\ref{ap:diag}. Indeed,
\be
\label{dK}
\frac{\pd K(z'^i,z^j)}{\pd q_\beta} = \frac{\pd }{\pd q_\beta}{\mathcal{G}_2^{ij}(z',z|{\bf q})\over\mathcal{G}_0({\bf q})} dz'^i dz^j =
\\
= \left({ \frac{\pd }{\pd q_\beta}\mathcal{G}_2^{ij}(z',z|{\bf q})\over\mathcal{G}_0({\bf q})}-{\mathcal{G}_2^{ij}(z',z|{\bf q})\over\mathcal{G}_0({\bf q})}{ \frac{\pd }{\pd q_\beta}\mathcal{G}_0({\bf q})\over\mathcal{G}_0({\bf q})}\right)dz'^i dz^j
\ee
as follows from \rf{berg} for the conformal block with two currents inserted $\mathcal{G}_2^{ij}(z',z|{\bf q}) = \mathcal{G}_2^{ij}(z',z|{\bf q})_0=\langle J_i(z')J_j(z)\mathcal{O}(\mathbf{q}) \rangle_0$ when projected to the vanishing $a$-periods \rf{aper} or the charges in
the intermediate channels (note, that the Bergman tau-function does not depend on these charges). Proceeding
with \rf{dK} and using $\frac{\d}{\d q_\beta}=L_{-1}^\beta$ one gets therefore
\be
\label{dKc}
\frac{\pd K(z'^i,z^j)}{\pd q_\beta} =  \left({\langle J_i(z')J_j(z)L_{-1}^\beta\mathcal{O}(\mathbf{q}) \rangle_0\over\langle \mathcal{O}(\mathbf{q}) \rangle_0 }-{\langle J_i(z')J_j(z)\mathcal{O}(\mathbf{q}) \rangle_0\over\langle \mathcal{O}(\mathbf{q}) \rangle_0}{ \langle L_{-1}^\beta\mathcal{O}(\mathbf{q}) \rangle_0\over\langle \mathcal{O}(\mathbf{q}) \rangle_0}\right)dz'^i dz^j
\ee
where we have used the obvious notations
\be
\label{notR}
\langle \mathcal{O}(\mathbf{q}) \rangle_0 = \langle \prod_{\alpha=1}^{2L}\mathcal{O}_\alpha(q_\alpha) \rangle_0 =
\langle\mc{O}_{s_1}(q_1)\mc{O}_{s_1^{-1}}(q_2)...\mc{O}_{s_L}(q_{2L-1})\mc{O}_{s_L^{-1}}(q_{2L})\rangle_0
\\
\langle L_{-1}^\beta\mathcal{O}(\mathbf{q}) \rangle_0 = \langle \left(L_{-1}\mathcal{O}_\beta(q_\beta)\right)\prod_{\alpha\neq\beta}\mathcal{O}_\alpha(q_\alpha) \rangle_0
= \half\oint_{q_\beta}\sum_k d\zeta \langle :J_k^2(\zeta):\mathcal{O}(\mathbf{q}) \rangle_0
\\
\langle J_i(z')J_j(z)L_{-1}^\beta\mathcal{O}(\mathbf{q}) \rangle_0 = \half\oint_{q_\beta}\sum_k d\zeta \langle J_i(z')J_j(z):J_k^2(\zeta):\mathcal{O}(\mathbf{q}) \rangle_0
\ee
where the integration $\oint_{q_\beta} d\zeta$ is performed on the base $\mathbb{P}^1$. Applying now in the r.h.s. the Wick theorem (see Appendix~\ref{ap:diag} for details), one gets
\be
\label{wickrauch}
\half\langle J_i(z')J_j(z):J_k^2(\zeta):\mathcal{O}(\mathbf{q}) \rangle_0 \langle \mathcal{O}(\mathbf{q}) \rangle_0 =
\half\langle J_i(z')J_j(z) \mathcal{O}(\mathbf{q}) \rangle_0\langle :J_k^2(\zeta):\mathcal{O}(\mathbf{q}) \rangle_0 +
\\
+ \langle J_i(z')J_k(\zeta)\mathcal{O}(\mathbf{q}) \rangle_0 \langle J_j(z)J_k(\zeta)\mathcal{O}(\mathbf{q}) \rangle_0
\ee
which means for \rf{dKc}, that
\be
\frac{\pd K(z'^i,z^j)}{\pd q_\beta} = \oint_{q_\beta}\sum_k d\zeta
{\langle J_i(z')J_k(\zeta)\mathcal{O}(\mathbf{q}) \rangle_0\over\langle \mathcal{O}(\mathbf{q}) \rangle_0} { \langle J_j(z)J_k(\zeta)\mathcal{O}(\mathbf{q}) \rangle_0\over\langle \mathcal{O}(\mathbf{q}) \rangle_0}dz'^i dz^j =
\\
= \oint_{q_\beta}\sum_k \frac{K(z'^i,\zeta^k)K(z^j,\zeta^k)}{d\zeta}=\Sum_{\pi(P)=q_\beta}\underset{P}{\Res}\frac{K(z'^i,P)K(z^j,P)}{d z(P)}
\ee
where we have used that $\oint_{q_\beta}\Sum_k = \Sum_{\pi(P)=q_\beta}\underset{P}{\Res}$. Hence, the same methods, which give rise to explicit formula for the main part $\tau_{SW}(\mathbf{a},\mathbf{q})$ of the
exact conformal block, ensure also the consistency of definition of the quasiclassical correction $\tau_B(\bs q)$.

\section{Isomonodromic $\tau$-function \label{ss:isotau}}

The full exact conformal block equals therefore
\be
\mc{G}_0({\bf q}|{\bf a})=\tau_B({\bf q})\exp\left({\half \Sum_{IJ}a_I\mc{T}_{IJ}({\bf q}) a_J +\Sum_I a_I U_I({\bf q},{\bf r})+\half Q({\bf r})}\right)
\label{Gex}
\ee
According to \cite{Painleve,PG} the $\tau$-functions of the isomonodromy problem \cite{SJM}
on sphere with four marked points $0,q,1,\infty$
can be decomposed into a linear combination of the corresponding conformal blocks~\footnote{This relation has been predicted in \cite{Knizhnik}, see also \cite{Novikov}
for a slightly different observation of the same kind.}.
This expansion looks as
\be
\tau_{IM}(q)=\Sum_{\bs w\in \mathcal{Q}(\mathfrak{sl}_N)} e^{(\bs b,\bs w)}
C^{(0q)}_{\bs w}(\bs\theta_0,\bs\theta_q,\bs a,\mu_{0q},\nu_{0q})
C_{\bs w}^{(1\infty)}(\bs\theta_1,\bs\theta_\infty,\bs a,\mu_{1\infty},\nu_{1\infty})\times\\\times
q^{\frac12(\bs\sigma_{0t}+\bs w,\bs\sigma_{0t}+\bs w)-\frac12(\bs\theta_0,\bs\theta_0)-\frac12(\bs\theta_t,\bs\theta_t)}\mathcal{B}_{\bs w}(
\{\bs\theta_i\},\bs a,\mu_{0q},\nu_{0q},\mu_{1\infty},\nu_{1\infty};q)
\label{imcft}
\ee
and can be tested, both numerically and exactly for some degenerate values of the W-charges $\bs\theta$ of the fields \cite{PG,GIL}. In \rf{imcft} the normalization of conformal block $\mathcal{B}_{\bs w}(\bullet;q)$ is chosen to be $\mathcal{B}_{\bs w}(\bullet;q) = 1 + O(q)$ and $C^{(\bullet)}_{\bs w}(\bullet)$ as usually denote the corresponding 3-point structure constants (all these quantities
in the case of $W(\mathfrak{sl}_N)=W_N$ blocks with $N>2$ depend on extra parameters $\{\mu,\nu\}$, being the coordinates on the moduli space of flat connections on 3-punctured sphere, and for their generic
values the conformal blocks $\mathcal{B}_{\bs w}(\bullet;q)$ are not defined algebraically, see \cite{PG} for more details).

We now conjecture that such decomposition exists also for conformal blocks considered above. Moreover, then a natural guess is, that the
structure constants have such a form that
\be
C^{(0q)}_{\bs w}(\bs\theta_0,\bs\theta_q,\bs a,\mu_{0q},\nu_{0q})
C_{\bs w}^{(1\infty)}(\bs\theta_1,\bs\theta_\infty,\bs a,\mu_{1\infty},\nu_{1\infty})
q^{\frac12(\bs a+\bs w,\bs a+\bs w)-\frac12(\bs\theta_0,\bs\theta_0)-\frac12(\bs\theta_q,\bs\theta_q)}
\cdot
\\
\cdot\mathcal{B}_{\bs w}(
\{\bs\theta_i\},\bs a,\mu_{0q},\nu_{0q},\mu_{1\infty},\nu_{1\infty};q)=
\mc{G}_0(\{\bs\theta_i\},\bs a+\bs w;q)
\label{CCBG}
\ee
i.e. they are absorbed into our definition of the W-block of the twist fields, and this can be
extended from four to arbitrary number of even $2L$ points on sphere. This conjecture can be
easily checked in the $N=2$ case, where the structure constants for the values, corresponding to the Picard solution \cite{Painleve,ILT},  coincide exactly with given by degenerate period matrices in
\rf{Gex}, when applied to the case of the Zamolodchikov conformal blocks \cite{Quiver} (see sect.~\ref{ss:examples} and Appendix~\ref{app:matrix}).

It means that in order to get isomonodromic $\tau$-functions from the exact conformal blocks \rf{Gex}
one has just to sum up the series (for the arbitrary number of points one has to replace the root lattice of $\mathcal{Q}(\mathfrak{sl}_N)=\mathbb{Z}^{N-1}$ by the lattice $\mathbb{Z}^g$, where $g=g(\mathcal{C})$ is the genus of the cover)
\be
\tau_{IM}({\bf q}|\bs a,\bs b)=\Sum_{\bs n\in\mathbb{Z}^g}\mathcal{G}_0({\bs q}|\bs a+\bs n)e^{(\bs n,\bs b)}=
\tau_B(\bs q)\exp\left(\half Q(\mathbf{r})\right)\times\\\times
\Sum_{\bs n\in \mathbb{Z}^g}\exp\left(\frac12 \left({\bf a+n},\mc{T}{\bf(a+n)}\right)+ (\bs U,\bs a+\bs n)+(\bs b,\bs n)\right)=\\=
\tau_B(\bs q)\exp\left(\half Q(\mathbf{r})\right)\Theta\left[\begin{matrix}\bs a\\\bs b\end{matrix}\right]\left(\bs U\right)
\label{tauKor}
\ee
which is easily expressed through the theta-function. One gets in this way
exactly the Korotkin isomonodromic $\tau$-function, where the only difference of this expression with proposed in \cite[formula 6.10]{Korotkin} is in the term $Q(\bs r)$, which is not expressed globally through the coordinates of the branch points in the case of non-abelian monodromy group.
This fact supports both our conjectures: about the form of the structure constants, and about the general correspondence between the isomonodromic deformations and conformal field theory.

Formula \rf{tauKor} has also clear meaning in the context of gauge theory/topological string correspondence. It has been noticed yet in \cite{NO}, that the CFT free fermion representation
exists only for the dual partition function, which is obtained from the gauge-theory matrix
element (conformal block) by a Fourier transform~\footnote{The fact, that only the Fourier-Legendre transformed
quantity can be identified with partition function in string theory has been established recently
in quite general context from their transformation properties in \cite{deWit}.}. We plan to return to this issue separately in the context of the free fermion representation for the exact W-conformal
blocks.

\section{Examples \label{ss:examples}}

There are several well-known examples of the conformal blocks corresponding to Abelian monodromy groups. All of them basically come from the Zamolodchikov exact conformal
block \cite[formula 3.29]{ZamAT} for the Ashkin-Teller model, defined on the families of hyperelliptic curves
\be
y^2=\prod_{\alpha=1}^{2L}(z-q_\alpha)
\ee
with projection $\pi:(y,z)\mapsto z$. Parameters $\mathbf{r}$ are absent here, so the result is
just $\mc{G}_0({\bf q})=\tau_B({\bf q})\exp\left(\frac12 \Sum_{IJ}a_I\mc{T}_{IJ}({\bf q}) a_J\right)$,
where for the hyperelliptic period matrices one gets from \rf{SWR1} the well-known Rauch formulas
(see e.g. \cite{Quiver} and references therein).

When the hyperelliptic curve degenerates (see Appendix~\ref{app:matrix}), this formula gives
\be
\mc{G}_0({\bf q})\approx 4^{-\sum a_I^2-
\left(\sum a_I\right)^2}\Prod_{I=1}^g(q_{2I}-q_{2I-1})^{a_I^2-\frac18}\Prod_{I>J}^g(q_{2I}-q_{2J})^{2a_Ia_J} R^{-\left(\sum a_I\right)^2}
\approx\\\approx 4^{-\sum a_I^2-
\left(\sum a_I\right)^2}\cdot\Prod_{I=1}^g\epsilon_I^{a_I^2-\frac18}R^{-\left(\sum a_I\right)^2}\cdot\Prod_{I>J}^g(q_{2I}-q_{2J})^{2a_Ia_J}
\label{Gdeg}
\ee
Here in the r.h.s. the second factor comes from the OPE \rf{ope}, i.e.
$\mc{O}(q_{2I}-\epsilon_I)\mc{O}(q_{2I})\sim \epsilon_I^{a_I^2-\frac18}V_{a_I}(q_{2I})+\ldots$, while the third
one is just the correlator $\langle\prod V_{a_I}(q_{2I})\rangle$. Hence, the first most important factor corresponds to the non-trivial product of the structure constants in \rf{CCBG}, which acquires here
a very simple form.
The main point of this observation is that normalization of \rf{Gex} automatically contains not only $q^\#$ factors, but also the structure constants,
and we have already exploited such conjecture for general situation in sect.~\ref{ss:isotau}, since the argument with degenerate tau-function can be easily extended.

These observations have an obvious generalization for the $\mathbb{Z}_N$-curves
\be
y^N=\prod_{\alpha=1}^{2L}(z-q_\alpha)^{k_\alpha}
\ee
with the same projection $\pi:(y,z)\mapsto z$. The main contribution to the answer $\tau_{SW}=\exp\left(\frac12 \Sum_{IJ}a_I\mc{T}_{IJ}({\bf q}) a_J\right)$ comes just from a general reasoning as in sect.~\ref{ss:twists} and to make it more explicit one can use the Rauch formulas
for $\mathbb{Z}_N$-curves, which express everything in terms of the coordinates $\{q\}$ on the projection, since there is no summing over preimages in formulas like \rf{SWR1}.

Let us now turn to an elementary new example with non-abelian monodromy group.
Notice, first, that a simple genus $g(\mathcal{C})=0$ curve
\be
\label{cur}
y^3 = x^2(1-x)
\ee
gives rise to the curve with non-abelian monodromy group if one takes a different (from $\mathbb{Z}_3$-option $\pi_x:(y,x)\mapsto x$) projection $\pi_y:(y,x)\mapsto y$. For the curve $\mathcal{C}$, which is just
a sphere or $\mathbb{P}^1$ itself, one gets here two essentially different (and unrelated!) setups, corresponding to differently chosen functions $x$ or $y$.

In the first case our construction leads, for example, to the formulas
\be
\label{Txx}
\langle T(x) \rangle_\mathcal{O} = {\langle T(x)\mathcal{O}_{s}(0)\mathcal{O}_{s^{-1}}(1)\rangle\over \langle\mathcal{O}_{s}(0)\mathcal{O}_{s^{-1}}(1)\rangle}  =
\frac14\{ \xi; x\} = {1\over 9x^2(x-1)^2}
\ee
where $x={1\over 1+ \xi^3}$ in terms of the global coordinate $\xi$ on $\mathcal{C}$, and
this formula fixes the insertions at $x=0,1$ to be
the twist operators
for $s=(123)$, with $\Delta(s) = {l^2-1\over 24l}\ \stackreb{l=3}{=}\ {1\over 9}$.

However, for a similar correlator on $y$-sphere
\be
\label{Tyy}
\langle T(y) \rangle_{\tilde{\mathcal{O}}} = {\langle T(y)\prod_{A = 0,1,2,3}{\tilde{\mathcal{O}}}(y_A)\rangle\over \langle\prod_{A = 0,1,2,3}{\tilde{\mathcal{O}}}(y_A)\rangle} = {1+54y^3\over (27y^3-4)^2y^2} =
\\
 =\sum_{A = 0,1,2,3} \left({1\over 16(y-y_A)^2} + {u_A\over y-y_A}\right)
\\
y_0=u_0=0,\ \ \ \ \ \ 3y_k=-8u_k=2^{2/3}e^{2\pi i(k-1)/3},\ k=1,2,3
\ee
one has to insert the twist operators for $\tilde{s}=(12)(3)$ of dimension $\Delta(\tilde{s}) = {\tilde{l}^2-1\over 24\tilde{l}}\ \stackreb{\tilde{l}=2}{=}\ {1\over 16}$. The r.h.s. here follows from
summation of
\be
{1\over 12}\{ \xi; y\} = {\xi(\xi^3+4)(1+\xi^3)^4\over 2(2\xi^3-1)^4} = {\xi^5(3y+\xi)\over 2y(2\xi-3y)^4}
\label{Schy}
\ee
where
\be
\label{uni}
y={\xi\over 1+ \xi^3},\ \ \ \ \xi\in \mathcal{C}=\mathbb{P}^1
\ee
To get \rf{Tyy} one has to sum \rf{Schy}
over $\pi(\xi)=y$, or three solutions of the equation $R(\xi) = \xi^3-{1\over y} \xi +1 = 0$, i.e.
\be
\label{STy}
\langle T(y) \rangle_{\mc C} = {1\over 12}\sum_\beta\{ \xi^{(\beta)}; y\} =
\sum_\beta \res_{\xi=\xi^{(\beta)}}\left({\xi^5(3y+\xi)\over 2y(2\xi-3y)^4}d\log R(\xi)\right) =
\\
= - {1\over 2y}\left(\res_{\xi=3y/2}+\res_{\xi=\infty}\right)\left({\xi^5(3y+\xi)R'(\xi)
\over (2\xi-3y)^4R(\xi)}d\xi\right) = {1+54y^3\over (27y^3-4)^2y^2}
\ee
in contrast to the sum over three sheets of the cover $\pi(\xi)=x$, which gives only a factor $\langle T(x) \rangle_\mathcal{O} = 3\cdot \{ \xi; x\}/12$.

To analyze the simplest nontrivial $\tau$-function for non-abelian monodromy group, let us consider the deformation of the
formula from \rf{uni}
for $z=1/y=\xi^2+1/\xi$, i.e. the
cover $\pi: \mathcal{C}=\mathbb{P}^1_\xi\rightarrow\mathbb{P}^1_z$
given by 1-parametric family
\be
z={(2\xi-t^2+1)^2(\xi-4)\over (t-3)^2(t^2-2t-3)\xi}
\ee
The parametrization is adjusted in the way that the branching points $dz=0$ are at
\be
\xi=\half (t^2-1),\ z=0
\\
\xi=1+t,\ z=1
\\
\xi=1-t,\ z=q(t)={(t+3)^3(t-1)\over (t-3)^3(t+1)}
\ee
together with $\xi=\infty$, $z=\infty$.

One also has non-branching points above the branched ones $\xi=4$, $z=0$; $\xi=(t-1)^2$, $z=1$; $\xi=(t+1)^2$, $z=q(t)$.
Now we rewrite these points in our notation
\be
\begin{array}{lll}
\xi_0^1=4,\quad&\xi_0^2=\frac12(t^2-1),\quad&\xi_0^3=\frac12(t^2-1)\\
\xi_q^1=(t+1)^2,\quad&\xi_q^2=1-t,\quad&\xi_q^3=1-t\\
\xi_1^1=(t-1)^2,\quad&\xi_1^2=1+t,\quad&\xi_1^3=1+t\\
\xi_\infty^1=0,\quad&\xi_\infty^2=\infty,\quad&\xi_\infty^3=\infty
\end{array}
\ee
Using an explicit formula \rf{Qfin0} and the definition \rf{KK} of $\tau_B$ one can write down the result for the $\tau$-function
\be
\tau(t)=\tau_B(t)\exp\left(\frac12Q_\mathbf{r}(t)\right) = \\
= (t-3)^{\delta_3-\frac13}(t-1)^{\delta_1-\frac18}t^{\delta_0+\frac1{24}}(t+1)^{
\delta_{-1}}(t+3)^{\delta_{-3}+\frac1{24}}
\label{taualg}
\ee
where $\delta_\nu=\delta_\nu(\mathbf{r})$ are given by some particular quadratic forms
\be
\delta_3=9r_q^2-9r_\infty^2\\
\delta_1=r_0^2-4r_0r_1+4r_1^2+8r_0r_q-4r_1r_q+r_q^2-4r_0r_\infty+8r_1r_\infty-4r_qr_\infty+4r_\infty^2\\
\delta_0=-9r_1^2-9r_q^2\\
\delta_{-1}=4 r_0^2+8r_0r_1+4r_1^2-4 r_0r_q+7r_q^2-4 r_0r_\infty-4r_1r_\infty+8r_qr_\infty+r_\infty^2\\
\delta_{-3}=-9r_0^2-9r_q^2
\ee
while their ``semiclassical'' shifts come from the Bergman $\tau$-function. Notice that isomonodromic
function \rf{taualg} looks very similar to the tau-functions of algebraic solutions of the Painlev\'e
VI equation \cite[examples 5-7]{Painleve}, but depends on essentially more parameters.

An interesting, but yet unclear observation is that in this
example $\tau_B(t)$ itself can be represented as
\be
\tau_B(t)=\exp\left(\frac12 Q({\bf \tilde r})\right)
\ee
for several particular choices of parameters ${\bf\tilde r}$, e.g.
\be
(\tilde r_0,\tilde r_q,\tilde r_1,\tilde r_\infty)=(\frac{\sqrt{7}}{12\sqrt{3}},
-\frac{\sqrt{7}}{12\sqrt{3}},\frac{i}{4\sqrt{3}},\frac{\sqrt{7}}{12\sqrt{3}})\\
(\tilde r_0,\tilde r_q,\tilde r_1,\tilde r_\infty)=(\frac{i}{12\sqrt{3}},\frac{i}{12\sqrt{3}},\frac{i}{12\sqrt{3}},\frac{\sqrt{5}}{12})
\ee
whereas all other (altogether eight) solutions are obtained after the action of the Galois group generated by $\sqrt{3}\mapsto -\sqrt{3}$,
$\sqrt{5}\mapsto -\sqrt{5}$ and $i\mapsto-i$. Notice that this statement is nevertheless nontrivial because we express five variables $\delta_i$ in terms of only four variables $\tilde r_i$.

\section{Conclusions}

We have presented above an explicit construction of the conformal blocks of the twist fields in the conformal theory with integer central charges and extended W-symmetry. We have computed the W-charges
of these twist fields and show that their Verma modules are non-degenerate from the point of view of W-algebra representation theory. The obtained exact formulas for the corresponding conformal blocks were derived intensively using the correspondence between two-dimensional conformal and four-dimensional supersymmetric gauge theory.
We also checked that so constructed exact conformal blocks, when considered in the context of isomonodromy/CFT correspondence, give rise to the
isomonodromic $\tau$-functions of the quasipermutation type.

We believe that it is only the beginning of the story and, finally, would like to present a list (certainly not complete) of unresolved yet problems. For the conformal field theory side these obviously include:
\begin{itemize}
\item What is the algebraic structure of the W-algebra representations corresponding to the twist-field vertex operators, and in particular -- what are the form-factors or matrix elements of these operators?
\item Already for the twist fields representations the analysis of this paper should be supplemented by study of the W-analogs of the higher-twist representations \cite{ApiZam} and of the W-representations
    at ``dual values'' of the central charges (an example of such block for the Virasoro case can be found
    in \cite{ZamD}).
\item Finally, perhaps the most intriguing question is -- what is the constructive generalization of these vertex operators to non-exactly-solvable case?
\end{itemize}

However, the main intriguing part still corresponds to the side of supersymmetric gauge theory, where the resolution of these problems can help to understand their properties in the ``unavoidable'' regime of strong coupling, where even the Lagrangian formulation is not known. We are going to return to these questions elsewhere.

\section*{Appendix}

\appendix

\section{Diagram technique\label{ap:diag}}

In order to compute the correlators of the currents \rf{Rcor} the first useful observation is that one can embed $\mathfrak{sl}_N\subset \mathfrak{gl}_N$ and introduce an extra current $h(z)$, commuting with
$J_i(z)$, such that
\be
h(z)h(z')=\frac{1/N}{(z-z')^2}+reg.,\ \ \ \ \
h(z)\mc{O}(q)=reg.
\ee
Introduce the $\mathfrak{gl}_N$ currents
\be
\tilde J_i(z)=J_i(z)+h(z)
\ee
which satisfy the OPE
\be
\tilde J_i(z)\tilde J_k(z')=\frac{\delta_{jk}}{(z-z')^2}+reg.
\ee
and their normally-ordered averages are the same as for $J_i(z)$ since
\be
\langle :J_{i_1}(z_1)\ldots J_{i_m}(z_m)h(z_{m+1})\ldots h(z_n):\rangle_{\mc O}=\\=
\langle :J_{i_1}(z_1)\ldots J_{i_m}(z_m):\rangle_\mc{O}\cdot\langle :h(z_{m+1})\ldots h(z_n):\rangle=0
\ee
Hence, one can simply to replace $J_i(z)\rightarrow \tilde J_i(z)$ in \rf{Rcor}, so below we just compute the averages for the $\mathfrak{gl}_N$ currents.

The normal ordering for two currents at colliding points is given by
\be
:J_{i}(z)J_j(z'):dz\,dz'=J_i(z)J_j(z')dz\,dz'-\frac{\delta_{ij}dz\,dz'}{(z-z')^2}=
\\
= J_i(z)J_j(z')dz\,dz'-\delta_{ij}K_0(z,z')
\label{noB}
\ee
i.e. it is defined by subtracting the canonical meromorphic bidifferential on the \emph{base curve}, since it corresponds to the
\emph{vacuum} expectation value of the Gaussian fields.  Normal ordering for the correlators of many currents is defined, as usual, by the Wick theorem.

Similarly to \rf{correlators} consider now
\be
\langle J_{i_1}(z_1):J_{i_2}(z_2)\ldots J_{i_n}(z_n):\rangle_{\mc O}\ dz_1\ldots dz_n=
\\
= dS( z_1^{i_1})
\langle :J_{i_2}(z_2)\ldots J_{i_n}(z_n):\rangle_{\mc O}\ dz_2\ldots dz_n+\\+
\Sum_{j=2}^n K( z_1^{i_1},z_j^{i_j})\langle:J_{i_2}(z_2)\ldots\widehat{J_{i_j}(z_j)}\ldots J_{i_n}(z_n):\rangle_{\mc O}\ dz_2\ldots
\widehat{dz_j}\ldots dz_n
\ee
where by $ z_k^i=\pi^{-1}(z_k)^i$ we have denoted the preimages on the cover. This formula is again obtained just from the analytic structure of this expression as 1-form in the first variable. The next formula comes from the application of the Wick theorem and \rf{noB}
\be
\langle J_{i_1}(z_1):J_{i_2}(z_2)\ldots J_{i_n}(z_n):\rangle_{\mc O}dz_1\ldots dz_n=
\langle :J_{i_1}(z_1)\ldots J_{i_n}(z_n):\rangle_{\mc O}dz_1\ldots dz_n+\\+
\Sum_{j=2}^n\delta_{ij}K_0(z_1,z_j)\langle:J_{i_2}(z_2)\ldots\widehat{J_{i_j}(z_j)}\ldots J_{i_n}(z_n):\rangle_{\mc O}dz_2\ldots
\widehat{dz_j}\ldots dz_n
\ee
Subtracting them, one gets the recurrence relation
\be
\langle :J_{i_1}(z_1)\ldots J_{i_n}(z_n):\rangle_{\mc O}dz_1\ldots dz_n=dS( z_1^{i_1})
\langle :J_{i_2}(z_2)\ldots J_{i_n}(z_n):\rangle_{\mc O}dz_2\ldots dz_n+\\+
\Sum_{j=2}^n \hat K_{{i_1}{i_j}}(z_1,z_j)\langle:J_{i_2}(z_2)\ldots\widehat{J_{i_j}(z_j)}
\ldots J_{i_n}(z_n):\rangle_{\mc O}dz_2\ldots\widehat{dz_j}\ldots dz_n
\ee
where we have introduced the ``propagator''
\be
\hat K_{ij}(z_1,z_2)=K( z_1^i, z_2^j)-\delta_{ij}K_0(z_1,z_2)
\ee
Graphically for the result this recurrence produces one can write

\begin{picture}(100,130)
\put(0,50){
\put(110,57){
\put(0,0){$\mc{R}_{i}(z_1)=$}\put(62,3){\circle*{5}}\put(68,3){$i$}\put(82,0){$=dS( z_1^i)$}
}

\put(50,17){
\put(0,0){$\mc{R}_{ij}(z_1,z_2)=$}\multiput(82,13)(0,-20){2}{\circle*{5}}\put(88,13){$i$}\put(88,-7){$j$}
\put(107,0){$+$}
\put(50,0){\multiput(82,13)(0,-20){2}{\circle*{5}}\put(88,13){$i$}\put(88,-7){$j$}}\put(132,-9){\line(0,1){20}}

\put(152,0){$=dS( z_1^i)dS( z_2^j)+\hat K_{ij}( z_1^i, z_2^j)$}
}}

\put(10,17){
\put(0,0){$\mc{R}_{ijk}(z_1,z_2,z_3)=$}

\multiput(0,0)(70,0){4}{
\multiput(102,12)(0,-20){2}{\circle*{5}}\put(94,12){$i$}\put(94,-9){$j$}
\put(122,3){\circle*{5}}\put(127,0){$k$}}
\multiput(143,0)(70,0){3}{$+$}

\put(172,-10){\line(0,1){20}}
\put(241,-9){\line(5,3){19}}
\put(311,13){\line(2,-1){20}}\put(350,0){$=$}
}

\end{picture}

$$=dS( z_1^i)dS( z_2^j)dS( z_3^k)+\hat K_{ij}(z_1,z_2)dS( z_3^k)+\hat K_{jk}(z_2,z_3)dS( z_1^i)+
\hat K_{ik}(z_1,z_3)dS( z_2^j)$$
These expressions have very simple meaning: the full correlation function is expressed through
the only possible connected parts $\mc{R}^c$, which are
$\mc{R}^c_i(z_1)=dS( z_1^i)$, $\mc{R}^c_{ij}(z_1,z_2)=\hat K_{ij}(z_1,z_2)$, while $\mc{R}_{ijk}^c(z_1,z_2,z_3)=0$ and all higher connected parts vanish. The so constructed four
point functions $\mc{R}_{ijkl}(z_1,z_2,z_3,z_4)$ at coinciding arguments (and at least pairwise
coinciding labels of the sheets of the cover) were used in sect.~\ref{ss:higher} for computation of the higher W-charges.

\section{ $W_4(z)$ and the primary field\label{ap:primarity}}

Here we study the OPE of $W_4(z)$ with $T(z)$ and show an explicit correction which makes this field primary.
\be
W_4(z)=\Sum_{ijkl}C^{ijkl}:J_i(z)J_j(z)J_k(z)J_l(z):
\ee
where $C^{ijkl}$ is completely symmetric tensor, $C^{ijkl}=\frac1{24}$ when $i\neq j\neq k\neq l$ and $C^{ijkl}=0$ otherwise.
\be
T(z)W_4(z')=\frac6{(z-z')^4}\Sum_{ijkl}\left(\delta_{ij}-\frac1N\right)C^{ijkl}:J_k(z)J_l(z):+\\
+\frac{4 W_4(z')}{(z-z')^2}+
\frac{\pd W_4(z')}{z-z'}+reg.
\ee
The first sum equals
\be
6\Sum_{ij}\left(\delta_{ij}-\frac1N\right)C_{ijkl}=-\frac{(N-2)(N-3)}{4N}(1-\delta_{ij})
\ee
Using now the fact that $\Sum_i J_i(z)=0$ we get
\be
\label{TWan}
T(z)W_4(z')=\frac{(N-2)(N-3)}{8N}\frac{T(z')}{(z-z')^4}+
 \frac{4 W_4(z')}{(z-z')^2}+\frac{\pd W_4(z')}{z-z'}+reg.
\ee
There is also another well-known field $\Lambda(z)=(TT)(z)-\frac3{10}\pd^2 T(z)$,
where $(TT)(z)=\oint_z\frac{dw}{w-z}T(w)T(z)$, with the OPE
\be
T(z)\Lambda(z')=\left(c+\frac{22}5\right)\frac{T(z')}{(z-z')^4}+\frac{4\Lambda(z')}{(z-z')^2}+
\frac{\pd\Lambda(z')}{z-z'}+reg.
\ee
One can therefore cancel an anomalous term in \rf{TWan} just
introducing
\be
\tilde W_4(z)=W_4(z)-\frac{(N-2)(N-3)}{8(N+\frac{17}{5})}\Lambda(z)
\ee
which is already a primary conformal field. Its charge therefore is given by the formula
\be
\tilde w_4=w_4-\frac{(N-2)(N-3)}{8(5N+17)}\Delta(5\Delta+1)
\ee

\section{Degenerate period matrix\label{app:matrix}}

\begin{figure}
\centering
\includegraphics[width=10cm]{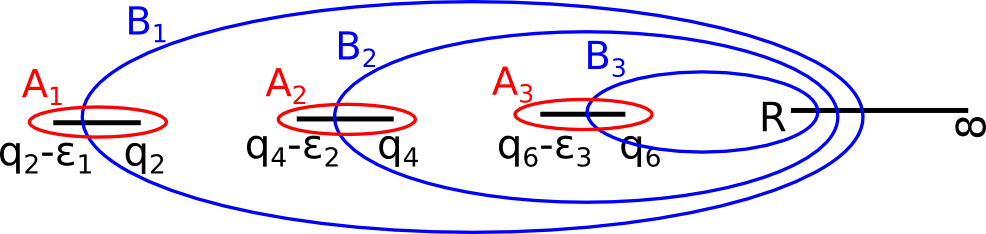}
\caption{Degenerate hyperelliptic curve with chosen basis in $H_1$.}
\label{fi:hypell}
\end{figure}

Here we compute the period matrix of the genus $g$ hyperelliptic curve (see fig.~\ref{fi:hypell})
\be
y^2=(z-R)\prod_{I=1}^g(z-q_{2I})(z-q_{2I}+\epsilon_I)=(z-R)\prod_{I=1}^g(z-q_{2I})(z-q_{2I-1})\label{hypercurve}
\ee
in the degenerate limit $\epsilon_I\to 0$, $R\to\infty$  up to the terms of order $O(\epsilon_I)$ and $O\left(\frac1R\right)$
(this equivalence will be denoted by ``$\approx$''). The normalized first kind Abelian differentials
with such accuracy are
\be
d\omega_I=\sqrt{q_{2I}-R}\prod_{K\neq I}(z-q_{2K})\frac{dz}{y},\ \ \ \
\frac1{2\pi i}\Oint_{A_J}d\omega_I\approx\delta_{IJ}
\ee
since $\frac{z-q_I}{\sqrt{(z-q_I)(z-q_I+\epsilon_I)}}\approx 1$ when $z$ goes far from $q_I$.
First we compute the off-diagonal matrix element $\mc{T}_{IJ}$ for $J>I$
\be
\mc{T}_{IJ}=\Oint_{B_J}d\omega_I\approx -2\Int_{q_J}^R\sqrt{\frac{q_{2I}-R}{z-R}}\frac{dz}{z-q_{2I}}\approx-2\log 4+
2\log\frac{q_{2J}-q_{2I}}{R}
\ee
and then a little bit more complicated diagonal element
\be
\mc{T}_{II}=\Oint_{B_I}d\omega_I\approx-2\Int_{q_{2I}}^R\sqrt\frac{q_{2I}-R}{z-R}\frac{dz}{\sqrt{(z-q_{2I})
(z-q_{2I}+\epsilon_I)}}\approx\\\approx
2\Int_{q_{2I}}^R\left(1-\sqrt\frac{q_{2I}-R}{z-R}\right)\frac{dz}{z-q_{2I}}-2
\Int_{q_{2I}}^R\frac{dz}{\sqrt{(z-q_{2I})(z-q_{2I}+\epsilon_I)}}=\\=
-2\log 4-2\log 4+2\log\frac{\epsilon_I}{R}
\ee
where we have used the fact, that for our purposes in the expressions $\frac{f(z)}{\sqrt{(z-q_{2I})(z-q_{2I}+\epsilon_I)}}$ one can drop $\epsilon_I$ if $f(q_{2I})=0$.

Now using \rf{Gex} we can compute in this limit
\be
\tau_{SW}=\exp\left(\frac12\Sum_{I<J}a_I\mc{T}_{IJ}a_J\right)\approx\\\approx
\cdot 4^{-\sum a_I^2-
\left(\sum a_I\right)^2}\Prod_{I=1}^g(q_{2I}-q_{2I-1})^{a_I^2}\Prod_{I>J}^g(q_{2I}-q_{2J})^{2a_Ia_J} R^{-\left(\sum a_I\right)^2}
\ee
The result for $\tau_B({\bf q})$ in this simple hyperelliptic example can be taken from \cite{ZamAT}
\be
\tau_B({\bf q})=\Prod_{i<j}^{2g+1}(q_i-q_j)^{-\frac18}\times\left[\det_{IJ}\frac1{2\pi i}\Oint_{A_I}\frac{z^{J-1}dz}{y}\right]^{-\frac12}
\ee
where the determinant can be easily computed using \rf{hypercurve}
\be
\det_{IJ}\frac1{2\pi i}\Oint_{A_I}\frac{z^{J-1}dz}{y}\approx\det_{IJ}\frac{q_{2I}^{J-1}}{\sqrt{R}\Prod_{J \neq I}(q_{2I}-q_{2J})}=R^{-\frac g2}\Prod_{I>J}
(q_I-q_J)^{-1}
\ee
Altogether this gives the formula \rf{Gdeg} for the degenerate form of the hyperelliptic
Zamolodchikov exact conformal block.

\subsection*{Acknowledgements}

We would like to thank the KdV Institute of the University of Amsterdam, where this work has been almost completed, and our colleagues there, especially Gerard Helminck, for the warm hospitality. These results
have been preliminary reported at the workshop \emph{Geometric Invariants and Spectral Curves}, Leiden, June 2015, and we would like to thank its organizers for illuminating discussions there.
The work was supported by the joint Ukrainian-Russian (NASU-RFBR) project 01-01-14 (P.G.) and 14-01-90405 (A.M.), the work of P.G. has been also supported by joint NASU-CNRS project F14-2015, while the work of A.M. was also supported by RFBR-15-01-99504, by joint RFBR/JSPS project 15-51-50034 and by the Program of Support of Scientific Schools (NSh-1500.2014.2).
The paper was prepared within the framework of a subsidy granted to the National Research University Higher School of Economics by the Government of the Russian Federation for the implementation of the Global Competitiveness Program.


\begin{thebibliography}{99}
\bibitem{ZamW}
A.~Zamolodchikov,
Theor. Math. Phys, {\bf 65}:3, (1985), 1205–1213.


\bibitem{FZ}V.~Fateev and A.~Zamolodchikov,
Nucl. Phys. {\bf B280}, (1987), 644-660.

\bibitem{FL}V.~Fateev and S.~Lukyanov,
Int. J. Mod. Phys. {\bf A3} (2), (1988), 507-520.


\bibitem{Wblocks}P.~Bowcock and G.M.T.~Watts,
Theor.~Math.~Phys. {\bf 98}, (1994), 350-356
[arXiv:hep-th/9309146].

\bibitem{LMN}
A.~S.~Losev, A.~Marshakov and N.~Nekrasov, in Ian Kogan memorial volume
{\it From fields to strings: circumnavigating theoretical physics},
581-621; [hep-th/0302191].

\bibitem{NO}
N.~Nekrasov and A.~Okounkov,
[hep-th/0306238].

\bibitem{AGT}
 L.~F.~Alday, D.~Gaiotto and Y.~Tachikawa,
  Lett.\ Math.\ Phys.\  {\bf 91} (2010) 167
  [arXiv:0906.3219 [hep-th]].

\bibitem{Nek}
N.~Nekrasov,
  Adv.\ Theor.\ Math.\ Phys.\  {\bf 7} (2004) 831
  [arXiv:hep-th/0206161];\\
 N.~Nekrasov and V.~Pestun,
  [arXiv:1211.2240 [hep-th]].


\bibitem{SW}N.~Seiberg and E.~Witten
Nucl. Phys. {\bf B426}, 19, (1994),
[arXiv:hep-th/9407087].

\bibitem{KriW}
I.~Krichever,
Commun. Pure. Appl. Math. {\bf 47} (1992) 437
[arXiv: hep-th/9205110].

\bibitem{AMtau}
  A.~Marshakov,
  JHEP {\bf 1307} (2013) 068,
  [arXiv:1303.0753 [hep-th]].

\bibitem{Quiver}P.~Gavrylenko and A.~Marshakov,
JHEP {\bf 0514}, (2014), 097,  [arXiv:1312.6382 [hep-th]].

\bibitem{BPZ} A.~Belavin, A.~Polyakov and A.~Zamolodchikov,
Nucl. Phys. {\bf B241}, (1984), 333-380.

\bibitem{ZamD}
Al.~Zamolodchikov,
JETP {\bf 90}, (1986), 1808-1818.

\bibitem{ZamAT}
Al.~Zamolodchikov, Nucl. Phys. \textbf{B285}, [FS19], (1987), 481-503.

\bibitem{ApiZam}
S.~Apikyan and Al.~Zamolodchikov,
JETP {\bf 92}, (1987), 34-45.

\bibitem{Painleve}
  O.~Gamayun, N.~Iorgov and O.~Lisovyy,
  JHEP,  {\bf 1210}, (2012), 38,
  [hep-th/1207.0787].

\bibitem{Litv1}
V.~Fateev and A.~Litvinov,
JHEP {\bf 0711}, (2007), 002,
[arXiv:0709.3806 [hep-th]].

\bibitem{SJM} M.~Sato, T.~Miwa and M.~Jimbo,
Publ. RIMS Kyoto Univ. {\bf 14},
(1978), 223–267; {\bf 15}, (1979), 201–278; {\bf 15}, (1979), 577–629;{\bf 15},
(1979), 871–972; {\bf 16}, (1980),
531–584.

\bibitem{PG}
P.~Gavrylenko ,
JHEP {\bf 0915}, (2015), 167,
[arXiv:1505.00259 [hep-th]].

\bibitem{Knizhnik}V.~Knizhnik,
Comm. Math. Phys. {\bf 112}, 4, (1987), 567-590;
Russian Physics Uspekhi, {\bf 159}, (1989), 401–453.

\bibitem{Novikov}
D.~Novikov,
Theor. Math. Phys. {\bf 161}, (2009), 1485–1496

\bibitem{BR} M.~Bershadsky and A.~Radul,
Int. J. Mod. Phys. {\bf A02}, (1987), 165;
Comm. Math. Phys. {\bf 116}, 4, (1988), 689-700.


\bibitem{orbi}
L.~Dixon, D.~Friedan, E.~Martinec and S.~Shenker,
Nucl. Phys. \textbf{B282}, (1987) 13-73.

\bibitem{Dubrovin}
B.~Dubrovin
Russ. Math. Surv. {\bf 36}, (1981), 11.

\bibitem{Mumford}
D.Mumford, {\sl Tata Lectures on Theta}, 1988.


\bibitem{Fay}
J.~Fay, {\sl Theta-functions on Riemann surfaces,}
Lect. Notes Math. {\bf 352}, Springer, N.Y. 1973.


\bibitem{Bilal}A.~Bilal,
Phys. Lett. {\bf B227}, 3–4, (1989), 406–410.

\bibitem{MM}
A.~Marshakov and A.~Morozov,
Nucl. Phys. {\bf B339}, 1, (1990), 79–94.

\bibitem{FLitv}
V.~Fateev and A.~Litvinov,
JHEP {\bf 1201}, (2012), 051,
[arXiv:1109.4042 [hep-th]].

\bibitem{KotKor1}
A.~Kokotov and D.~Korotkin,
Math. Phys., Anal. and Geom., {\bf 7}, (2004), 1, 47-96,
[arXiv:math-ph/0202034].


\bibitem{GIL}P.~Gavrylenko, N.~Iorgov and O.~Lisovyy,
to appear.

\bibitem{Korotkin}
D.~Korotkin, Math. Ann. {\bf 329}, 2, (2004), 335-364,
[arXiv:math-ph/0306061].

\bibitem{KotKor2}
A.~Kokotov and D.~Korotkin,
Int. Math. Res. Not. (2006), 1-34,
[arXiv:math-ph/0310008].

\bibitem{Fay1} Fay, John D.,
Memoirs of AMS, 1992, v.96, n. 464.

\bibitem{deWit}
  G.~L.~Cardoso, B.~de Wit and S.~Mahapatra,
  JHEP {\bf 1409} (2014) 096
  [arXiv:1406.5478 [hep-th]].

\bibitem{ILT} N.~Iorgov, O.~Lisovyy, J.~Teschner,
Comm.~Math.~Phys. {\bf 336}, (2015), 671-694
[arXiv:1401.6104~[hep-th]]


\end{thebibliography}
\end{document}